\documentclass[%
 aip,pof
 amsmath,amssymb,
 preprint,%
]{revtex4-1}

\usepackage{natbib}
\usepackage{graphicx}
\usepackage[toc,page]{appendix}
\usepackage{multirow}
\usepackage{mathbbol}
\usepackage{amssymb}
\usepackage{eqnarray,amsmath}
\usepackage{bm}
\usepackage{pifont}
\usepackage{epsfig}
\usepackage{psfrag}
\usepackage{epstopdf}
\usepackage{wasysym}
\usepackage[usenames]{color}
\usepackage[symbol]{footmisc}
\usepackage{xparse}
\usepackage{hyperref}
\usepackage[nameinlink,noabbrev]{cleveref}
\usepackage{pgfplots}
\usepackage{tikz}

\usetikzlibrary{shapes.misc,calc,patterns,
                decorations.pathmorphing,
                decorations.markings}

\tikzset{cross/.style={cross out, draw=black, minimum size=2*(#1-\pgflinewidth), 
         inner sep=0pt, outer sep=0pt},
cross/.default={1pt}}

\hypersetup{
 colorlinks = true, 
 urlcolor = black, 
 linkcolor = black, 
 citecolor = black 
 }

\graphicspath{{./}{./images/}}

\NewDocumentCommand\dsum{e{_^}}{{\displaystyle\sum_{#1}^{#2}}}
\DeclareMathOperator{\dint}{\displaystyle\int}

\begin{document}

\title{A fast and efficient tool to study the rheology of dense suspensions}

\author{Alessandro Monti}
\email[Corresponding author: ]{alessandro.monti@oist.jp}
\affiliation{Complex Fluids and Flows Unit, Okinawa Institute of Science and Technology \\
     1919-1 Tancha, Onna, Kunigami District, Okinawa 904-0495, Japan.}
\author{Vikram Rathee}
\affiliation{Micro/Bio/Nanofluidics Unit, Okinawa Institute of Science and Technology \\
     1919-1 Tancha, Onna, Kunigami District, Okinawa 904-0495, Japan.}
\author{Amy Q. Shen}
\affiliation{Micro/Bio/Nanofluidics Unit, Okinawa Institute of Science and Technology \\
     1919-1 Tancha, Onna, Kunigami District, Okinawa 904-0495, Japan.}
\author{Marco E. Rosti}
\email[Corresponding author: ]{marco.rosti@oist.jp}
\affiliation{Complex Fluids and Flows Unit, Okinawa Institute of Science and Technology \\
     1919-1 Tancha, Onna, Kunigami District, Okinawa 904-0495, Japan.}


\date{\today}

\begin{abstract}
    A cutting-edge software that adopts an optimised searching algorithm
    is presented to tackle the Newton-Euler equations 
    governing the dynamics of dense suspensions in Newtonian fluids. 
    In particular, we propose an implementation of a fixed-radius near 
    neighbours search based on an efficient counting sort algorithm 
    with an improved symmetric search. 
    The adopted search method drastically reduces the computational cost and 
    allows an efficient parallelisation even on a single node through the 
    multi-threading paradigm. Emphasis is also given to the memory efficiency of 
    the code since the history of the contacts among particles has to be traced
    to model the frictional contributions, when dealing with dense suspensions of 
    rheological interest that consider non-smooth interacting particles. 
    An effective procedure based on an estimate of the maximum number of the 
    smallest particles surrounding the largest one (given the radii distribution)
    and a sort applied only to the surrounding particles only is implemented, 
    allowing us to effectively tackle the rheology of non-monodispersed particles 
    with high size-ratio in large domains.
    Finally, we present validations and verification of the numerical procedure, 
    by comparing with previous simulations and experiments, and present new 
    software capabilities.
\end{abstract}

\pacs{}

\keywords{Rheology, Dense suspensions, Complex fluids, Search algorithm.}

\maketitle


\section{Introduction}
The physics of granular particles suspended in a fluid has become of 
fundamental importance in the last few decades since their heavy 
involvement in many natural and industrial processes.
For instance, clarifying the behaviour of blood flows is a high-priority
topic since it can potentially lead to the diagnosis and prevention of 
cardiovascular diseases \citep{TAKEISHI2019}.
From a geophysical viewpoint, phenomena like 
sediment transport are highly
attractive since dense suspensions contribute in transporting 
nutrients and providing ecological habitats while shaping the surrounding 
landscape \citep{VERCRUYSSE2017}. 
Industries working with paints or adhesives, instead, are interested in the control 
of the flow to guarantee the required performance and quality of the final products
approaching the market \citep{ELEY2019}.

The physics of the processes listed above is dominated by the 
interactions between the particles and fluid that determine the 
flowability properties of the granular suspensions.
Unscrambling the cumbersome microscopic mechanisms of the interactions between 
the particles shaping those materials and their flowing behaviour 
has become a high-priority topic, with a particular need for finding 
a physically meaningful constitutive law \citep{TANNER2018}. Exceptionally challenging
is the rheological behaviour shown by dense suspensions, i.e. suspensions with a
similar volume fraction of particles and liquid. In fact, under stress, dense suspensions 
may show rheological behaviours that span from a yielded behaviour to
discontinuous shear-thickening \citep{WYART2014,GUAZZELLI2018,LEE2021,SIVADASAN2019}.

Despite the practical relevance, a few analytical studies dealing only with very 
dilute \citep{EINSTEIN1905} and semi-dilute \citep{BATCHELOR1972} granular 
suspensions have been successfully carried out. The major mathematical difficulties
stand in the lack of proper analytical formulations that deal with multi-body 
interactions, especially for very large volume fractions, where enduring direct 
contacts become fundamental \citep{GUAZZELLI2018}. 
Conversely to theoretical approaches, numerical algorithms are ideal for tackling the 
numerous particles interacting with each other.
Numerical techniques, spanning from molecular dynamics \citep{ALDER1959,VERLET1967} 
to the solution of the linear Stokes equations for every suspended particle 
\citep{SIEROU2001}, stormed the field of granular suspensions.
Recently, since the work done in the seminal paper by Cundall and Strack \citep{CUNDALL1979},
the discrete (or distinct) element method (DEM) for studying dense suspensions and granular 
flows has become the standard for the flow analysis. 
Modern approaches to DEM reduce the description of the particle-fluid and 
particle-particle interactions to minimal models able to correctly address the
macroscopic behaviour of the flow without excessive loss of information. On this fashion,
simplified relationships for short-range interactions have been proposed, e.g.
for hydrodynamic interactions \citep{BALL1997,SETO2013} and for adhesive-repulsive
contributions (a satisfactory overview can be found in the works of 
\citep{ISRAELACHVILI2011,LI2011,SETO2013,SINGH2019}). Concerning contacts,
the most implemented technique is the soft-particle model, where contacts among 
several particles and adhesion effects are handled \citep{LI2011}. 
In particular, Luding \citep{LUDING2008} developed a minimal model that, based on 
the original work by Cundall and Strack \citep{CUNDALL1979}, addresses the surface 
roughness of the particles with tangential contacts replicating the sliding, 
rolling and torsion resistances, with the option of reproducing both static and 
dynamic frictions. More advanced contact models contemplating non-linearity, 
hysteresis and complex effects can be found in his work \citep{LUDING2008}. 
Many other models have been adopted in the field of dense suspensions and DEM 
\citep{TOWNSEND2017,SIVADASAN2019}, the interested reader is referred to the reviews 
by Li \textit{et al.} \citep{LI2011} and Guo and Curtis \citep{GUO2015} for a 
complete overview.

The major drawback of DEM is the expensive computation of 
short-range interactions acting on every particle. The computational cost, if care 
is not taken, may quadratically increase with the number of particles considered, 
limiting the size of the domain analysable or the length of the simulated process. 
To reduce the computational cost,
algorithms of fixed-radius near neighbours search \citep{BENTLEY1975} have
been adopted and become the standard in DEM implementations. 
A widely implemented method is related to the
use of Verlet lists \citep{VERLET1967}. The Verlet list is a 
data structure that stores and tracks all particles within a cut-off 
distance of each other. The list can be naively built by
checking the $N(N-1)/2$ distances (cost $O(N^2)$) between particles 
or, with more modern techniques, through efficient cell-lists or 
tree-structures that largely reduce the computational cost ($O(N)$
for cell-lists, $O(N\log N)$ for tree structures) \citep{HOWARD2016}. 
In particular, the former consist in dividing the computational box into 
a Cartesian lattice (with spacing $\Delta$ that accommodates the particles 
within a pre-selected cut-off distance) and assigning the particles to the 
points of the lattice. After that, for each particle a distance-checking 
procedure is applied to all the other particles belonging to the $27$ 
neighbouring cells.
This procedure, however, may bring to a high computational and memory 
cost when dealing with highly polydispersed particles, since the cut-off 
distance is usually calibrated with the largest particles; therefore,
a cell may contain a large number of small particles that must be 
enquired for neighbourhood. While computationally the extra-cost can be 
decreased by adopting a cut-off radius that suits the smallest particle
of the set, and extending the size of the searching-stencil for the 
larger particles as done in LAMMPS, memory-wise the cost will remain of
the order $O(N^2)$, since the cell-list has to spatially cover the whole
simulation box.

To further reduce the computational cost, the Verlet-lists are updated every
$n$ time steps, calibrated such that within the $n-1$ remaining steps
the locations of the neighbouring particles do not significantly vary
\citep{VERLET1967}. This last point, however, is very delicate since
updating the Verlet list too frequently is computationally expensive
and, on the contrary, increasing the time-period of the update can 
introduce errors in the computation.

Here, following the work by \citet{HOETZLEIN2014}, we implement a 
highly parallelisable method based on the cell-list approach that does 
not require the explicit computation of a Verlet list, thus zeroing its
memory-cost, since the particles will be simply reordered within the 
Cartesian lattice in a new array by means of a very efficient counting-sort 
algorithm (computational cost $O(N)$). Since the particles are ordered by 
the cell, it is straight-forward to look for neighbours within the adjacent 
cells. The counting-sort algorithm will be applied to the particles at every
time iteration, avoiding any loss of information on the neighbouring pairs. 
Within the framework of this manuscript, we propose and validate the new
software that combines the cutting-edge algorithm just mentioned for the neighbours 
query with methods that satisfactorily deal with complex
rheological behaviours observed in dense suspensions. For the latter problem,
a rigorous mathematical formulation will be provided.
Finally, the software is parallelised within the multi-threading architecture 
and is intended to be used without accessing the expensive computational power 
provided by super-computers or costly GPU accelerators. 

The manuscript is organised as follows. \Cref{sec:numMet} deals with the 
mathematical formulation, focusing on the models adopted and detailing the 
parameters that govern the flow. In addition, the numerical integration scheme
together with its implementation and parallel performance are described.
In \cref{sec:res},  we show the validation of the code by studying the rheology of dense 
suspensions to ensure the reliability of the results.
Some benchmark test on the potentiality of the code are also proposed.
Finally, in \cref{sec:concl} a summary of the code capabilities and novelty ends the manuscript.

\section{Numerical method}
\label{sec:numMet}
The in-house software used in this work models a dense, non-Brownian 
suspension of quasi-inertialess, neutrally-buoyant, quasi-rigid 
spherical particles in a shear-flow defined by the shear-rate, $\dot{\gamma}$, 
and the strain-rate tensor, 
$\mathbb{E}^\infty$ (with the only non-zero quantities,
$\mathbb{E}^\infty_{12}=\mathbb{E}^\infty_{21}=\dot{\gamma}/2$).
The code tackles the Newton-Euler equations
that govern the translational and rotational dynamics of the rigid 
particles,
\begin{equation}
    \begin{cases}
    m_i \dfrac{\mathrm{d}\boldsymbol{u}_i}{\mathrm{d}t}
    &=  \dsum^{}_{M} \boldsymbol{F}^M_{i}, \\
    \mathbb{I}_i \dfrac{\mathrm{d}\boldsymbol{\omega}_i}{\mathrm{d}t} + 
    \boldsymbol{\omega}_i \times (\mathbb{I}_i \boldsymbol{\omega}_i)
    &= \dsum^{}_{M} \boldsymbol{T}^M_{i},
    \end{cases}
    \label{eq:NE}
\end{equation}
where the subscript $i$ indicates the particle $i\in [1,N]$, being
$N$ the number of particles.
In \cref{eq:NE}, $\boldsymbol{F}_i^M$ and $\boldsymbol{T}_i^M$ denote
the force and torque applied to the centre of mass of the 
$i$th particle, with mass $m_i$ and inertia tensor $\mathbb{I}_i$. 
The translational and angular velocities are here denoted by the symbols
$\boldsymbol{u}_i$ and $\boldsymbol{\omega}_i$, respectively.
The superscript $M$ refers to the nature of the forces applied to System
\eqref{eq:NE}, resulting from particle-particle and 
particle-flow interactions. In particular, here we consider the
contributions arisen from hydrodynamics, inelastic contacts and 
electro-chemical effects. Specifically, the right-hand-side of 
\cref{eq:NE} can be written as
\begin{equation}
    \begin{cases}
    \dsum^{}_{M} \boldsymbol{F}^M_{i} &= 
    \dsum_{j=1}^{N_H} \boldsymbol{F}^H_{ij} +
    \dsum_{j=1}^{N_C} \boldsymbol{F}^C_{ij} +
    \dsum_{j=1}^{N_E} \boldsymbol{F}^E_{ij}, \\
    \dsum^{}_{M} \boldsymbol{T}^M_{i} &= 
    \dsum_{j=1}^{N_H} \boldsymbol{T}^H_{ij} +
    \dsum_{j=1}^{N_C} \boldsymbol{T}^C_{ij} +
    \dsum_{j=1}^{N_E} \boldsymbol{T}^E_{ij}, \\
    \end{cases}
    \label{eq:NErhs}
\end{equation}
where the summations are performed on the number of particles 
neighbouring the $i$th particle considered,  and
the superscripts $H$, $C$ and $E$ stand for hydrodynamics, inelastic 
contacts and electro-chemical effects, respectively.

\begin{figure}[htb]
    \centering
    \includegraphics[width=0.37\textwidth]{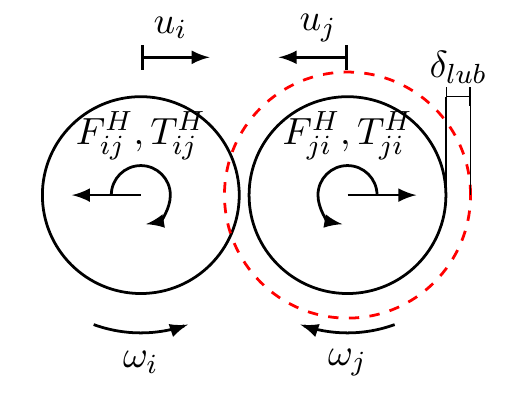}
%
%
    \caption{Sketch of the hydrodynamic interaction between the pair of 
             particles, $i,j$, having velocity and angular velocity $\boldsymbol{u}_i$,
             $\boldsymbol{u}_j$ and $\boldsymbol{\omega}_i$, $\boldsymbol{\omega}_j$,
             respectively. The hydrodynamics effect starts acting when the minimum
             surface distance between the particles is below $\delta_{lub}$.}
    \label{fig:lub}
\end{figure}

Concerning the hydrodynamics of the system, 
dense suspensions of rigid particles immersed in a low-Reynolds-number 
flow are subjected to a Stokes drag and a pair-wise, 
short-range lubrication force \citep{MARI2014}.
The latter is caused by the relative motion of particles that squeeze
the fluid flowing in the narrow gaps between them.
Following the work of Ball and Melrose \citep{BALL1997} and 
Mari \textit{et al.} \citep{MARI2014}, to replicate those contributions 
we implement a linear relationship between forces and velocities,
\begin{equation}
    \label{eq:hydro}
    \begin{pmatrix}
    \boldsymbol{F}^H_i \\
    \boldsymbol{F}^H_j \\
    \boldsymbol{T}^H_i \\
    \boldsymbol{T}^H_j
    \end{pmatrix}
    =-
    \begin{pmatrix}
    \mathbb{R}_{Stokes} + \mathbb{R}_{Lub}
    \end{pmatrix}_{ij}
    \cdot
    \begin{pmatrix}
    \boldsymbol{u}_i - \boldsymbol{U}^\infty(\boldsymbol{x}_i)\\
    \boldsymbol{u}_j - \boldsymbol{U}^\infty(\boldsymbol{x}_j)\\
    \boldsymbol{\omega}_i - \boldsymbol{\Omega}^\infty(\boldsymbol{x}_i)\\
    \boldsymbol{\omega}_j - \boldsymbol{\Omega}^\infty(\boldsymbol{x}_j)
    \end{pmatrix}
    +
    \begin{pmatrix}
    \mathbb{R}_{Lub}^{'(\mathbb{E})}\\
    \end{pmatrix}_{ij}
    \begin{pmatrix}
    \boldsymbol{n}\\
    \boldsymbol{n}\\
    \boldsymbol{n}_{\perp,i}\\
    \boldsymbol{n}_{\perp,j}
    \end{pmatrix},
\end{equation}
where $\boldsymbol{n} = (\boldsymbol{x}_j-\boldsymbol{x}_i)
/|\boldsymbol{x}_j-\boldsymbol{x}_i|$ is the centre-to-centre 
unit vector (being $\boldsymbol{x}_i$ the Cartesian position 
of the particle $i$), and $\boldsymbol{n}_{\perp,i}(A) = 
2\boldsymbol{n}\times [(A_i+A_i^T)\cdot \boldsymbol{n}]$
is a vector perpendicular to $\boldsymbol{n}$ (being $A_i$ the strain-rate 
tensor referred to the $i$th particle, $\mathbb{E}^\infty_i$, in 
this case), $\mathbb{R}_{Stokes}$ is a diagonal matrix that
models the Stokes drag forces and torques, including the Fax{\'e}n laws
for non-uniform shear-flows \citep{GUAZZELLI2011}, and 
$\mathbb{R}_{Lub}$ and $\mathbb{R}'_{Lub}$ are sparse
matrices 
that approximate the far-field mobility matrices, dropping the
low moment contributions and considering a pair resistance model
with only divergent terms \citep{BALL1997}. In particular, 
those sparse matrices model the contribution of the lubrication
through the squeeze, shear and pump modes \citep{MARI2014}.
The detailed formulation of the matrices defining the hydrodynamics
resistances are given in \cref{app:lub}. \Cref{fig:lub} sketches the 
lubrication forces and torques acting on the $i$th particle (left) 
due to the presence of the $j$th particle (right). The dashed red 
circumference highlights the region of influence of the lubricant 
interactions, parameterised with $\delta_{lub}$. Note that
beyond this threshold, i.e. even at mid-range distances between
the particles, the hydrodynamics is modelled only through the 
Stokes drag. We acknowledge that this is an approximation, however
in the context of dense suspensions it is acceptable since its
contribution is minimal.

\begin{figure}[htb]
    \centering
    \includegraphics[width=0.9\textwidth]{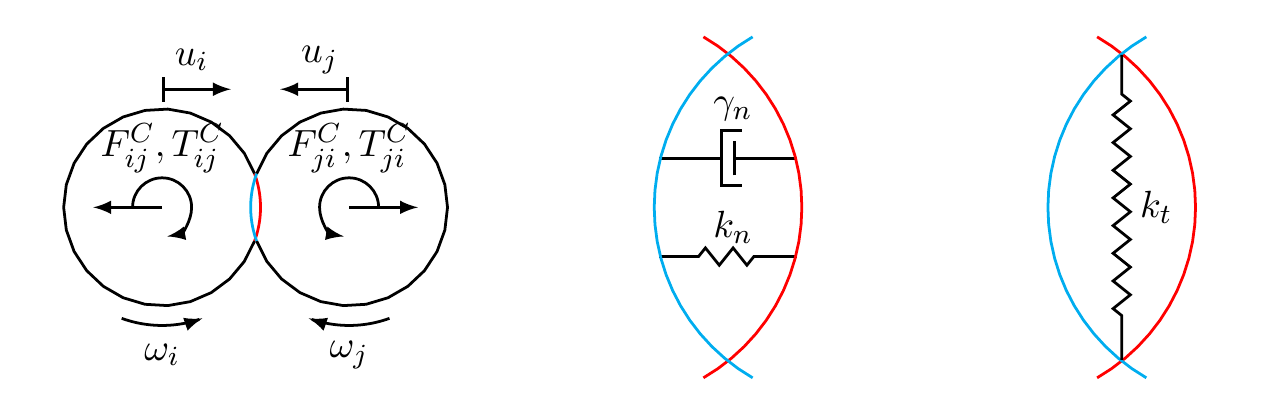}
    \caption{Sketch of the contacts model between the pair of 
             particles, $i,j$, having velocity and angular velocity $\boldsymbol{u}_i$,
             $\boldsymbol{u}_j$ and $\boldsymbol{\omega}_i$, $\boldsymbol{\omega}_j$,
             respectively. The spring-dashpot systems modelling the contacts start 
             acting when the surface distance, $d_{ij}$, between the particles is 
             negative (overlap).}
    \label{fig:contact}
\end{figure}

The contact contribution is modelled by the 
stick-and-slide method \citep{LUDING2008}, that is able to describe 
frictional contacts. This method mimics the inelastic contacts with 
spring-dashpot systems that start operating when the
surface distance between the spheres $i$ and $j$, $d_{ij} = 
|\boldsymbol{x}_i - \boldsymbol{x}_j|-(a_i+a_j)$ (being $a_i$ and
$a_j$ the particles radii), is negative (overlap).
Considering two particles, $i$ and $j$, the stick-and-slide model 
can be written as
\begin{subequations}
\begin{align}
    \label{eq:contact0}
    &\boldsymbol{F}^{C,nor}_{ij} = k_n d_{ij} \boldsymbol{n} 
                              + \gamma_n \boldsymbol{u}_{n,ij},\\
    &\boldsymbol{F}^{C,tan}_{ij} = k_t \boldsymbol{\xi}_{ij}, \\
    &\boldsymbol{F}^{C}_{ji} = -\boldsymbol{F}^{C}_{ij}, \\
    &\boldsymbol{T}^{C}_{ij} = a_i \boldsymbol{n} \times 
                            \boldsymbol{F}^{C,tan}_{ij}, \\
    &\boldsymbol{T}^{C}_{ji} = (a_j/a_i)\boldsymbol{T}^{C}_{ij},
\end{align}
\end{subequations}
where the Coulomb's law $|\boldsymbol{F}^{C,tan}_{ij}| \leq \mu_C
|\boldsymbol{F}^{C,nor}_{ij}|$ must be satisfied ($\mu_C$ being the 
friction coefficient). The superscripts $nor$ and $tan$ indicate 
the directions of the force (i.e. normal and tangential) and the 
parameters $k_n$, $k_t$ and $\gamma_n$ are the normal and tangential 
spring constants and the normal damping constant, respectively. 
The velocity $\boldsymbol{u}_{n,ij} = \boldsymbol{n} \otimes 
\boldsymbol{n} \cdot (\boldsymbol{u}_j-\boldsymbol{u}_i)$ is the normal 
velocity vector, while $\boldsymbol{\xi}_{ij}$ represents 
the stretch of the tangential spring. The latter is computed following the 
procedure described by Luding \citep{LUDING2008},
\begin{equation}
    \label{eq:csi}
    \boldsymbol{\xi}_{ij} = 
    \begin{cases}
        \mathbb{T}\dint_{t_0}^t\! \boldsymbol{u}_{t,ij}\, \mathrm{d}\tau, 
        & \text{if $|\boldsymbol{\xi}_{ij}|<
        \dfrac{\mu_C\left|\boldsymbol{F}^{C,nor}\right|}{k_t}$},\\[15pt]
        \dfrac{\mu_C\left|\boldsymbol{F}^{C,nor}\right|}{k_t} 
        \dfrac{\boldsymbol{\xi}_{ij}}
        {\left|\boldsymbol{\xi}_{ij}\right|}, 
        & \text{otherwise},
    \end{cases}
\end{equation}
where $\mathbb{T}=\mathbb{I}-\boldsymbol{n} \otimes \boldsymbol{n}$ is the
tangential projection tensor, $\boldsymbol{u}_{t,ij} =\mathbb{T} 
\cdot(\boldsymbol{u}_j-\boldsymbol{u}_i - (a_i \boldsymbol{\omega}_i + 
a_j\boldsymbol{\omega}_j)\times\boldsymbol{n})$ is the tangential velocity 
vector and $t_0$ is the time at the beginning of the collision between 
the particles $i$ and $j$ (with $\boldsymbol{\xi}_{ij}(t_0)=\boldsymbol{0}$).
A sketch of the model adopted in this work in shown in \cref{fig:contact}.
The possibility of adding the rolling resistance to the model, as in 
\citep{LUDING2008,SINGH2019}, is described in \ref{app:roll}.

\begin{figure}[htb]
    \centering
    \includegraphics[width=0.9\textwidth]{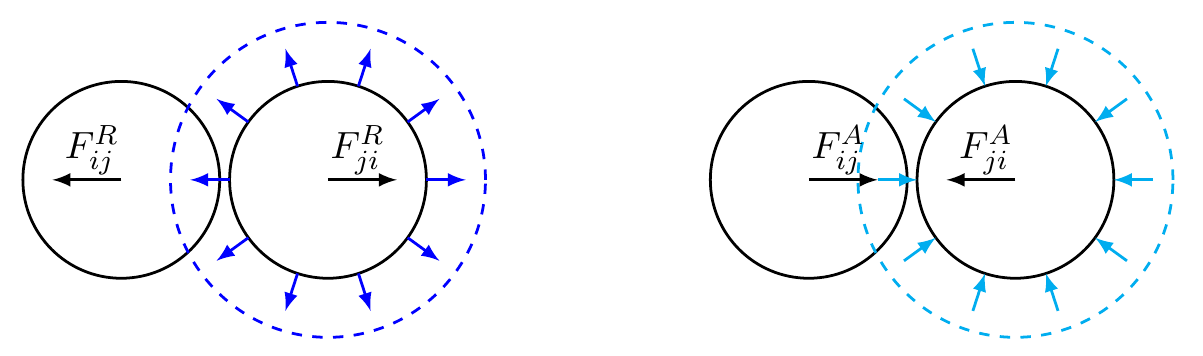}
    \caption{Sketch of the model adopted for the electro-chemical interaction 
             between the pair of particles, $i,j$. The repulsive interaction
             starts appearing when the minimum surface distance between the
             particles is below $20\lambda_D$, while the attractive contribution
             when the minimum surface distance is below $4 a_0$.}
    \label{fig:elch}
\end{figure}

The last contribution considered in this work are forces of conservative 
nature, that arise from the chemo-physical properties of the 
solid-fluid and solid-solid interactions and play a fundamental role 
in systems of dense suspensions \citep{GALVEZ2017,SINGH2019}.
In particular, the model, sketched in \cref{fig:elch}, addresses the
contribution by adding an inter-particle, distance-decaying repulsion force, 
\begin{subequations}\label{eq:rep}
\begin{align}
    &\boldsymbol{F}^R_{ij} =  
    \begin{cases}
        - F_{ER} \dfrac{\overline{a}}{a_{0}}, 
        e^{-d_{ij}/\lambda_D}\,\boldsymbol{n},
        & \text{if $d_{ij}\geq 0$},\\[15pt]
        - F_{ER} \dfrac{\overline{a}}{a_{0}}, 
        \,\boldsymbol{n},
        & \text{if $d_{ij} < 0$},
    \end{cases}\\
    &\boldsymbol{F}^R_{ji} = -\boldsymbol{F}^R_{ij},    
\end{align}
\end{subequations}
and an attraction force in van der Waals form,
\begin{subequations}\label{eq:att}
\begin{align}
    &\boldsymbol{F}^A_{ij} =  
    \begin{cases}
        \dfrac{A \overline{a}}{12(d_{ij}^2 + \epsilon^2)} 
        \,\boldsymbol{n} 
        & \text{if $d_{ij}\geq 0$},\\[15pt]
        \dfrac{A \overline{a}}{12\epsilon^2} 
        \,\boldsymbol{n},
        & \text{if $d_{ij} < 0$}.
    \end{cases}\\
    &\boldsymbol{F}^A_{ji} = -\boldsymbol{F}^A_{ij}.    
\end{align}
\end{subequations}
In \cref{eq:rep}, $F_{ER}$ is the module of the force, $a_0$ 
the reference radius, ${\overline{a}} = 2 a_i a_j/(a_i+a_j)$  
the harmonic radius and $\lambda_D$ is the Debye length that controls 
the decay of the force. In \cref{eq:att}, instead, $A$ is the 
Hamaker constant and $\epsilon$ is a small regularisation term 
(generally $\epsilon \approx 0.01\overline{a}$) introduced to avoid the 
singularity at contact, when $d_{ij}=0$.
To simplify the notation, below the repulsive and attractive effects may be
gathered together in the electro-chemical contribution,
$\boldsymbol{F}^E = \boldsymbol{F}^R+\boldsymbol{F}^A$, marked by the $E$
superscript.

The forcing contributions described above induce mechanical stress in the 
suspension that can be computed using the stresslets theory \citep{GUAZZELLI2011},
\begin{equation}
    \mathbb{\Sigma} = 2\eta_0 \mathbb{E}^\infty + 
    \dfrac{1}{V} \dsum_{i=1}^N \left( 
    \dsum_{j=1}^{N_H} \mathbb{S}^H_{ij} +
    \dsum_{j=1}^{N_C} \mathbb{S}^C_{ij} + 
    \dsum_{j=1}^{N_E} \mathbb{S}^E_{ij} \right),
    \label{eq:stress}
\end{equation}
where $\mathbb{\Sigma}$ is the bulk stress tensor, $\eta_0$ the 
viscosity of the carrier fluid and $V$ the volume of the simulated 
box. The stresslets due to the interactions, instead, are denoted by 
$\mathbb{S}$ and are defined as
\begin{subequations}
\begin{align}
    \mathbb{S}^S_{i} &= 20/3 \pi \eta_0 a_i^3 
      \left(1+a_i^2/10\nabla^2\right)\mathbb{E}^\infty_i\\
    \mathbb{S}^L_{ij} &= 0.5 \left[ 
    (\boldsymbol{x}_j-\boldsymbol{x}_i)\otimes \boldsymbol{F}^H_{ij} +  
    ((\boldsymbol{x}_j-\boldsymbol{x}_i)\otimes \boldsymbol{F}^H_{ij})^T
    \right], \\
    \mathbb{S}^C_{ij} &= 
    (\boldsymbol{x}_j-\boldsymbol{x}_i)\otimes \boldsymbol{F}^C_{ij}, \\
    \mathbb{S}^E_{ij} &= 
    (\boldsymbol{x}_j-\boldsymbol{x}_i)\otimes \boldsymbol{F}^E_{ij},
    \label{eq:stresslet}
\end{align}
\end{subequations}
where the hydrodynamic contribution ($H$) has been separated into the Stokes
($S$) and lubrication ($L$) terms, i.e. 
$\mathbb{S}^H_{ij}=\mathbb{S}^S_{i}+\mathbb{S}^L_{ij}$.
From the bulk stress tensor, $\mathbb{\Sigma}$, we can derive the
quantities of rheological interest, i.e. the shear-sress $\sigma$, 
the normal stress differences $N_1$ and $N_2$ and the pressure $\Pi$, 
defined as $\sigma=\mathbb{\Sigma}_{12}$, $N_1 = \mathbb{\Sigma}_{11}-
\mathbb{\Sigma}_{22}$, $N_2 = \mathbb{\Sigma}_{22}-\mathbb{\Sigma}_{33}$ 
and $\Pi = -Tr({\mathbb{\Sigma}})/3$. Normalising the shear-stress and the 
pressure with the properties of the carrier flow, we can define the
relative viscosity and the dimensionless pressure as $\eta_r=\sigma/
(\eta_0 \dot{\gamma})$ and $\eta_n=\Pi/(\eta_0\dot{\gamma})$, 
respectively.

Several parameters dominate the physics of the suspension described by 
System \eqref{eq:NE}. 
We consider the translational equation of \eqref{eq:NE},
\begin{equation}
    \dfrac{\rho a_0^3 U}{T}
    \dfrac{\mathrm{d}\boldsymbol{u}_i^*}{\mathrm{d}t^*}
    = \eta_0 a_0 U \dsum_{j=1}^{N_H}\boldsymbol{F}^{H,*}_{ij} 
    + k_{eq} a_0 \dsum_{j=1}^{N_C} \boldsymbol{F}^{C,*}_{ij}
    + F \dsum_{j=1}^{N_E} \boldsymbol{F}^{E,*}_{ij},
    \label{eq:ndNE}
\end{equation}
where we gathered the dimensional quantities (being the non-dimensional
quantities marked with the superscript $*$).
In \cref{eq:ndNE}, $U = a_0\dot{\gamma}$ is the typical velocity 
in a shear-flow, $T$ the time-scale introduced by the particles 
(in this case, the particles introduce no additional time-scales, 
therefore $T=1/\dot{\gamma}$), $k_{eq}$ the equivalent spring constant 
of the model represented in \cref{fig:contact} and $F$ the module 
of the force introduced by the electro-chemical interactions. 
Applying the Buckingham Pi theorem, the parameters are reduced to a 
set of non-dimensional numbers that can be tuned to outline the model 
desired. 
Considering the dimensional quantities in \cref{eq:ndNE},
$\{\rho,a_0,\eta_0,\dot{\gamma},k_{eq},F\}$ and choosing
$\{a_0,\eta_0,\dot{\gamma}\}$ as independent fundamental quantities, 
three non-dimensional groups emerge. The first one is the Stokes number,
\begin{equation}
    St = \dfrac{\rho a_0^2 \dot{\gamma}}{\eta_0} \ll 1,
    \label{eq:st}
\end{equation}
where the constraint $St \ll 1$ is applied to remain in the inertialess
regime. The non-dimensional stiffness, instead, governs
the importance of the contacts contributions compared to the hydrodynamic
term,
\begin{equation}
    \hat{k} = \dfrac{k_{n}}{\eta_0 a_0 \dot{\gamma}} \gg 1,
    \label{eq:stiff}
\end{equation}
where $\hat{k}\gg 1$ forces the particles to be rigid;
note that, the equivalent stiffness constant,
\begin{equation}
    k_{eq} = k_n \left( 1 + \dfrac{k_t}{k_n} + 
             \dfrac{\gamma_n \dot{\gamma}}{k_n} \right),
    \label{eq:keq}
\end{equation}
has been approximated to $k_n$ by adding additional constraints
$k_t \ll k_n$ and $\gamma_n\dot{\gamma}/k_n\ll 1$ \citep{MARI2014}.
Finally, the last non-dimensional group (the equivalent shear-rate) 
refers to the time-scale introduced by the electro-chemical 
contribution,
\begin{equation}
    \hat{\dot{\gamma}} = \dfrac{F(d_{ij}=0)}{\eta_0 a_0^2 \dot{\gamma}} 
                       = \dfrac{\dot{\gamma}_0}{\dot{\gamma}},
    \label{eq:ec}
\end{equation}
where we defined a new shear-rate, $\dot{\gamma}_0=F(d_{ij})/\eta_0 a_0^2$;
thus, tuning $F$ (i.e. adjusting $F_{ER}$ and the Hamaker constant) 
appropriately, a shear-rate dependency can be imposed to the suspension.

\subsection{Numerical algorithm}
Following the work by Ge \textit{et al.} \citep{GE2020,GE2020b}, 
the governing equations \eqref{eq:NE} are advanced in time with 
the modified velocity-Verlet explicit scheme \citep{GROOT1997},
\begin{subequations}
\begin{align}
    \boldsymbol{x}_i^{(n+1)} &= \boldsymbol{x}_i^{(n)} + 
    \Delta t \boldsymbol{u}_i^{(n)} +
    \dfrac{\Delta t^2}{2}\boldsymbol{\alpha}_i^{(n)},\\
    \boldsymbol{u}_i^{(n+1/2)} &= \boldsymbol{u}_i^{(n)} + 
    \dfrac{\Delta t}{2}\boldsymbol{\alpha}_i^{(n)}, \\
    \boldsymbol{\alpha}_i^{n+1} &= \boldsymbol{\mathcal{F}}
    \left(\boldsymbol{x}_i^{(n+1)},\boldsymbol{u}_i^{(n+1/2)}\right),\\
    \boldsymbol{u}_i^{(n+1)} &= \boldsymbol{u}_i^{(n)} + 
    \dfrac{\Delta t}{2}\left( \boldsymbol{\alpha}_i^{(n)} + 
    \boldsymbol{\alpha}_i^{(n+1)} \right),
\end{align}
\end{subequations}
where $\boldsymbol{x}_i^{(n)}$, $\boldsymbol{u}_i^{(n)}$ and 
$\boldsymbol{\alpha}_i^{(n)}$ denote the position (orientation), 
velocity (angular velocity) and acceleration (angular acceleration) 
vectors of the particle $i$, respectively, at time $t=n\Delta t$, and 
$\boldsymbol{\mathcal{F}}$ resembles the right hand side of \cref{eq:NE}.
The modified velocity-Verlet scheme is second-order accurate in time and, being
explicit, it requires a time-step $\Delta t$ that resolves the smallest
time-scale of the system (typically established by the stiffness of the
contacts; a good approximation of the time-step can be computed 
through the time constant that can be defined in the normal spring-dashpot system,
$\Delta t\dot{\gamma}=\gamma_n\dot{\gamma}/k_n$). 
Different from other approaches, e.g. considering an 
over-damped systems dropping the inertial terms \citep{MARI2014}, 
this approach allows us to avoid the inversion of the resistance matrix
of the lubrication forces.

The domain is a box of size $L_x\times L_y\times L_z$ (or $L^3$ if cubic)
with a shear-rate, $\dot{\gamma}$, applied along the $y$ 
direction. At the edges of the domain, the Lees-Edwards boundary 
conditions \citep{LEES1972} are adopted to remove the effect of 
the walls,
\begin{subequations}
    \label{eq:le}
    \begin{align}
        x &= 
        \begin{cases}
            (x+L_x-\dot{\gamma}L_y t) \mod L_x
            & \text{if $y > L_y$},\\
            (x+L_x+\dot{\gamma}L_y t) \mod L_x 
            & \text{if $y < 0$},\\
            (x+L_x) \mod L_x 
            & \text{otherwise},
        \end{cases}\\[10pt]
        y &= (y+L_y) \mod L_y,\\[10pt]
        z &= (z+L_z) \mod L_z,\\[10pt]
        u &= 
        \begin{cases}
            (u-\dot{\gamma}t) 
            & \text{if $y > L_y$},\\
            (u+\dot{\gamma}t) 
            & \text{if $y < 0$}.
        \end{cases}
    \end{align}
\end{subequations}

\begin{figure}[htb]
  \centering
  \includegraphics[width=0.9\textwidth]{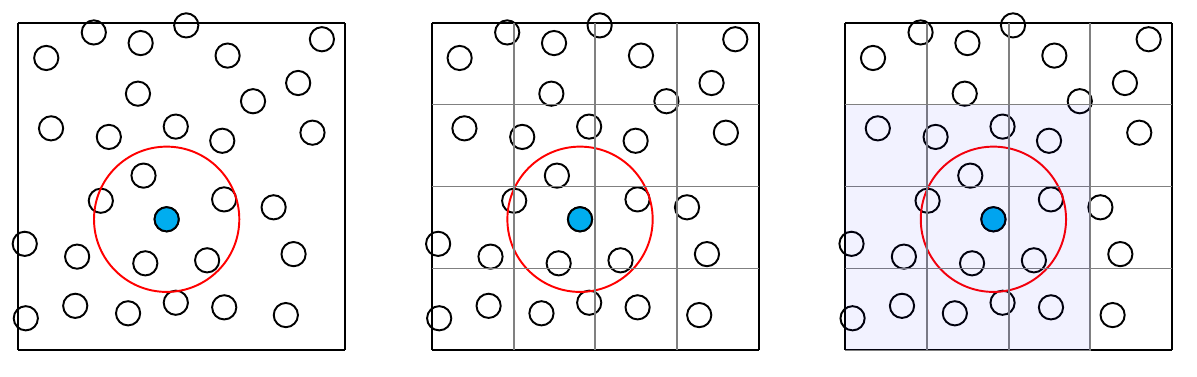}
  \caption{Sketch of the fixed-radius near-neighbours algorithm adopted 
           for finding the neighbours of the particle highlighted in blue, 
           within a distance $R$ marked by the red circle. 
           The domain (left panel) is subdivided, 
           first, into a coarse lattice (central panel) and the particles 
           are ordered via a counting sort algorithm. The search is 
           then carried out only on the $3\times 3$ neighbouring cells 
           (blue region, right panel).}
  \label{fig:alg}
\end{figure}

The interaction between the particles represents the most expensive part of
the algorithm from a computational point of view with a cost of $O(N^2)$, 
if a brute force algorithm is used.
To overcome the problem, we adopted a highly 
parallelisable, fixed-radius near neighbours algorithm 
\citep{BENTLEY1975,HOETZLEIN2014}.
The algorithm is based on rearranging the particles on a
coarser lattice (cell-list) by adopting an efficient counting
sort algorithm \citep{HOETZLEIN2014} that requires a computational 
cost $O(N)$ and few memory arrays of size $N$. The spacing of 
the cell-list $\Delta$ must suit the desired searching distance; 
a sketch of it is shown in \cref{fig:alg}.
By considering the near points of the coarser grid in the 
search for neighbours, i.e. the blue-filled region in 
\cref{fig:alg} (considering a $3\times 3$ macro searching-cell) 
the computational cost dramatically decreases to $O(cN)$, where 
$c \ll N$ is the number of particles inquired as neighbours and
depends on the lattice space $\Delta$.
The dependence of $c$ on $\Delta$ is related to the number of
particles that fit each macro-cell and an optimal choice of $\Delta$ 
is therefore related to the properties of the suspension.
For instance, fixing $\Delta = 2 a_{max}$, where $a_{max}$ is the 
radius of the largest particle in the suspension, minimizes the 
number of macro-cells to be inquired for the search of neighbours. 
This is ideal for monodisperse suspensions 
where there may be at maximum one particle per cell and may look reasonable
when considering polydispersed suspensions with not-very-high dispersion 
ratios. On the contrary, the choice may be detrimental for suspensions 
with $a_{min}\ll a_{max}$ ($a_{min}$ being the radius of the smallest particle), 
since it may result in the presence of a high number of particles within 
each macro-cell, with the limit of $c=N$ when $2a_{max}$ corresponds to the 
size of the domain. 
In this case, to reduce the number of particles within each searching cell,
a better choice is to set $\Delta<2 a_{max}$ and increase 
the stencil of neighbouring cells \citep{HOETZLEIN2014,HOWARD2016}.

\begin{figure}[htb]
  \centering
  \includegraphics[width=0.30\textwidth]{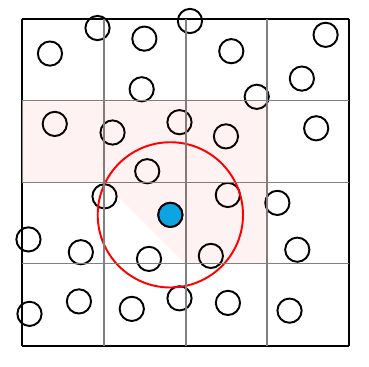}
  \caption{Symmetric fixed-radius near-neighbours algorithm. The search is carried out
           only within the red region. The initial random distribution of particles 
           within the domain guarantees the load-balance in case of parallelisation.}
  \label{fig:alg2}
\end{figure}

As a drawback, the search shown in \cref{fig:alg} implies that the 
forces and torques acting on the pair of particles $(i,j)$ are computed 
twice, although this procedure guarantees a better load balance 
through the processes when the code is parallelised. To avoid 
the double counting of the forces keeping a good load distribution 
among the processes, we propose a symmetric search, where the 
neighbours are scanned only over half of the points of the macro
searching-cell of the near Cartesian lattice, as sketched in 
\cref{fig:alg2}, thus obtaining an additional significant speed-up 
by halving the computational operations. Note that this procedure 
does not require any particular symmetry of the problem.

Memory-wise, instead, the main problem arises from the computation of 
the tangential stretch, \cref{eq:csi}, where the history of all contacts 
must be traced, with a potential memory cost $O(N^2)$. To reduce the cost, 
we use a reduced matrix with size $N\times M$, where $M$ is an overestimation 
of the maximum number of contacts per particle based on the volume of the spherical
shell surrounding the largest particle of the suspension, 
with thickness $\Delta a = 2 a_0$. 
The contacts, once established, are then recorded into a hash-map that charts them 
along with the indexes of the particles involved.
Then, every array of the map $(i,M)$ is sorted by the indexes and
the history of the contacts for the particle $i$ is sought over the sorted 
array $(i,M)$. This procedure has an actual memory cost $O(N)$, adding little
to the computational cost.

\begin{figure}[htb]
    \centering
    \includegraphics[width=0.5\textwidth]{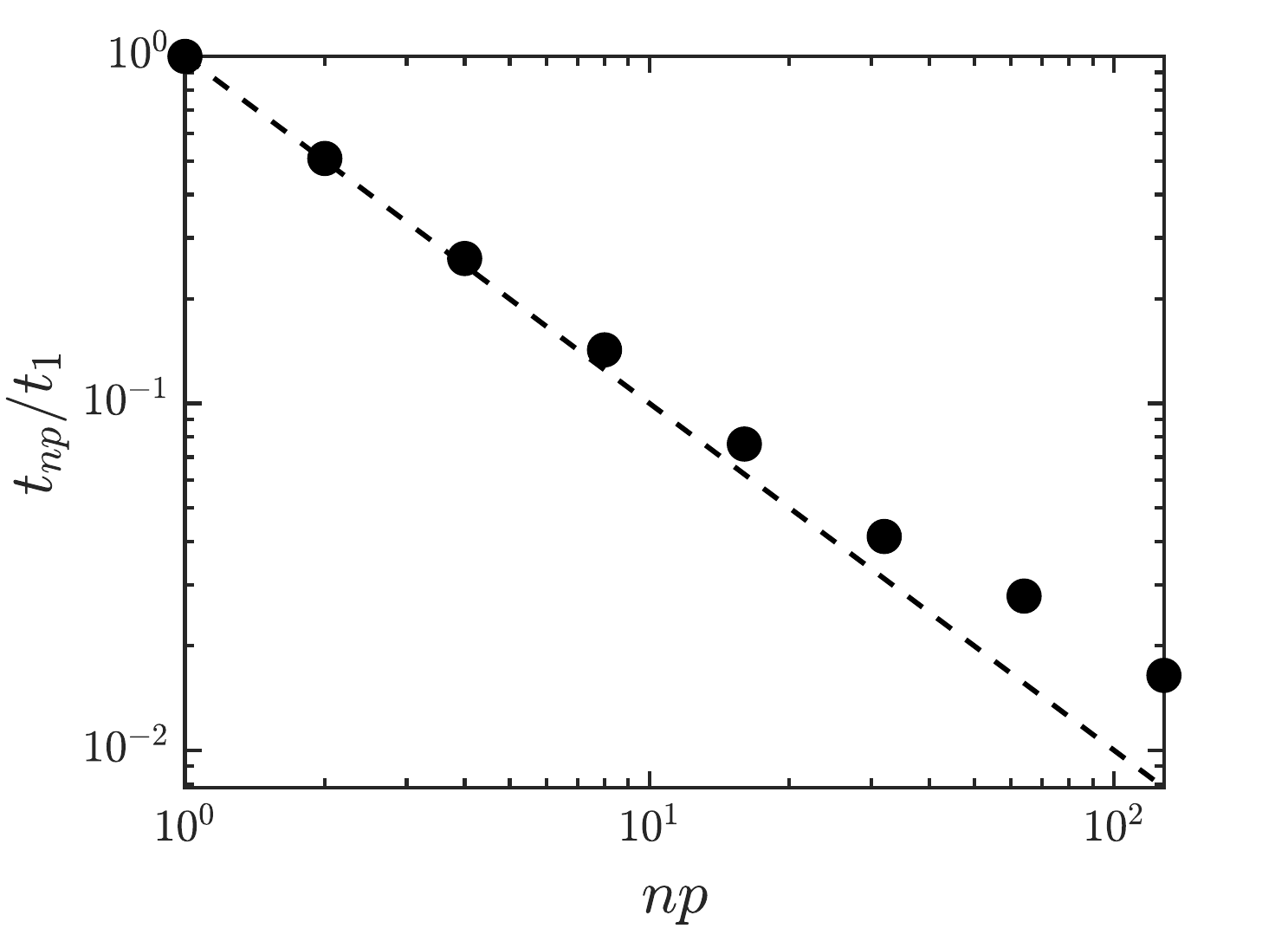}
    \caption{Scalability of the code. We considered a bidispersed
             suspension with number of particles $N=65536$ and volume fraction 
             $\phi=0.5$. The bidispersity of suspension is characterised by a
             bidispersion ratio $a_2/a_1=3$ and volume ratio $V_2/V_1=0.25$.
             The graph is obtained by increasing the number of processes $np$.
             The time spent per iteration by $np$ processes, $t_{np}$, 
             is normalised by the time spent by a single process, $t_1$.}
    \label{fig:scal}
\end{figure}

The parallelisation of the solver is reached by implementing the 
multi-threading paradigm on a single node with a multiprocessing system, 
using the OpenMP directives. The high-scalability of the code, obtained 
on a 2-socket AMD EPYC 7702 machine (zen2 microarchitecture), is shown in 
\cref{fig:scal}. The scalabiltiy graph is obtained by fixing the domain properties
and varying the number of processes used. The case considered is a bidispersed 
suspension with a high number of particles $N=65536$ to test also the memory-efficiency, 
volume fraction $\phi=0.5$, bidispersity ratio and volume $a_2/a_1=3$ and
$V_2/V_1=0.25$, respectively.
All the forcing contributions listed above in this section have been considered.
Referring to the configurations shown in \cref{fig:scal}, the total time 
per iteration (on average) in seconds is reported in \cref{tab:time}. The table 
also shows the breakdown (in percentage) of the total time needed by the several 
parts sampled within the core of the software. In particular, $t_1$ corresponds 
to the time needed to update the kinematical properties of the particles with 
the modified velocity-Verlet scheme, $t_2$ to sort the 
particles within the corresponding cells (counting sort), $t_3$ 
to compute the interactions between neighbours, and $t_4$ to
sort the hash-map of the frictional conctacts list, while
the amount of time needed for writing the outputs is negligible. 
\begin{table}[htb]
  \begin{center}
     \scalebox{0.8}{
     \setlength{\tabcolsep}{0.55em}
     \begin{tabular}{c|c|c|c|c|c|c|c|c}
         \hline
         \rule{0pt}{3ex}
         $np$ & $1$ & $2$ & $4$ 
         & $8$ & $16$ & $32$ & $64$ & $128$ \\[2mm]
         $t_{tot} [sec/iter]$ & $0.549$ & $0.279$ & $0.144$ 
         & $0.078$ & $0.042$ & $0.023$ & $0.015$ & $0.009$ \\[2mm]
         $t_{1} [\%]$ & $ 0.8$ & $ 0.8$ & $ 1.0$ & $ 2.1$ & $ 2.9$ & $ 5.5$ & $ 8.5$ & $ 8.6$ \\[2mm]
         $t_{2} [\%]$ & $ 0.2$ & $ 0.4$ & $ 0.5$ & $ 1.1$ & $ 1.4$ & $ 2.1$ & $ 3.2$ & $ 5.7$ \\[2mm]
         $t_{3} [\%]$ & $92.4$ & $92.5$ & $91.8$ & $88.0$ & $85.2$ & $79.0$ & $69.7$ & $70.1$ \\[2mm]
         $t_{4} [\%]$ & $ 6.6$ & $ 6.3$ & $ 6.7$ & $ 8.8$ & $10.5$ & $13.4$ & $18.6$ & $15.6$ \\\hline
     \end{tabular}}
 \caption{List of the average time per iteration needed to advance the particles varying the
      number of threads and its breakdown. In particular, $t_{tot}$ is the total time in seconds per iteration,
      $t_1$ is the percentage of time taken by the modified velocity-Verlet scheme, $t_2$ is the percentage of 
      time needed to sort the particles within the macro-cells (counting sort), $t_3$ is the percentage of
      time needed to compute the interactions between neighbouring particles and $t_4$ is the percentage of time
      needed for sorting the hash-map of the lists for the history of the frictional contacts.
      }
     \label{tab:time}
   \end{center}
\end{table}
The breakdown of the total time per iteration shows how the counting sort algorithm
is convenient for reordering the particles within the searching cells, with its cost not
exceeding the 6\% of the total time in the configuration considered.

For readers interested in the algorithm and in benchmarking the software,
the code is available at \url{https://github.com/marco-rosti/CFF-Ball-0x}.

\section{Results}
\label{sec:res}
In this section, the robustness and accuracy of the method proposed above for 
dense suspensions is investigated.
We consider, first, a relatively small domain with particles suspended in 
a uniform, plain shear-flow, and we compare the data obtained with 
studies available in the literature varying the volume fraction and the 
shear-rate; we then increase the domain size to test the memory-efficiency
of the software, and we study the behaviour of a dense suspension subject 
to a wavy shear-flow, the wave acting on the stream-shear cross-plane.
All initial conditions used in this work are generated with the sphere-packing 
software developed by Donev \citep{DONEV2006}.

\subsection{Uniform shear-flow}
The reliability of the code has been tested both in jamming transition scenarios
and in a shear-dependent behaviour of rheological relevance.

\begin{figure}[htb]
    \centering
    \includegraphics[width=0.5\textwidth]{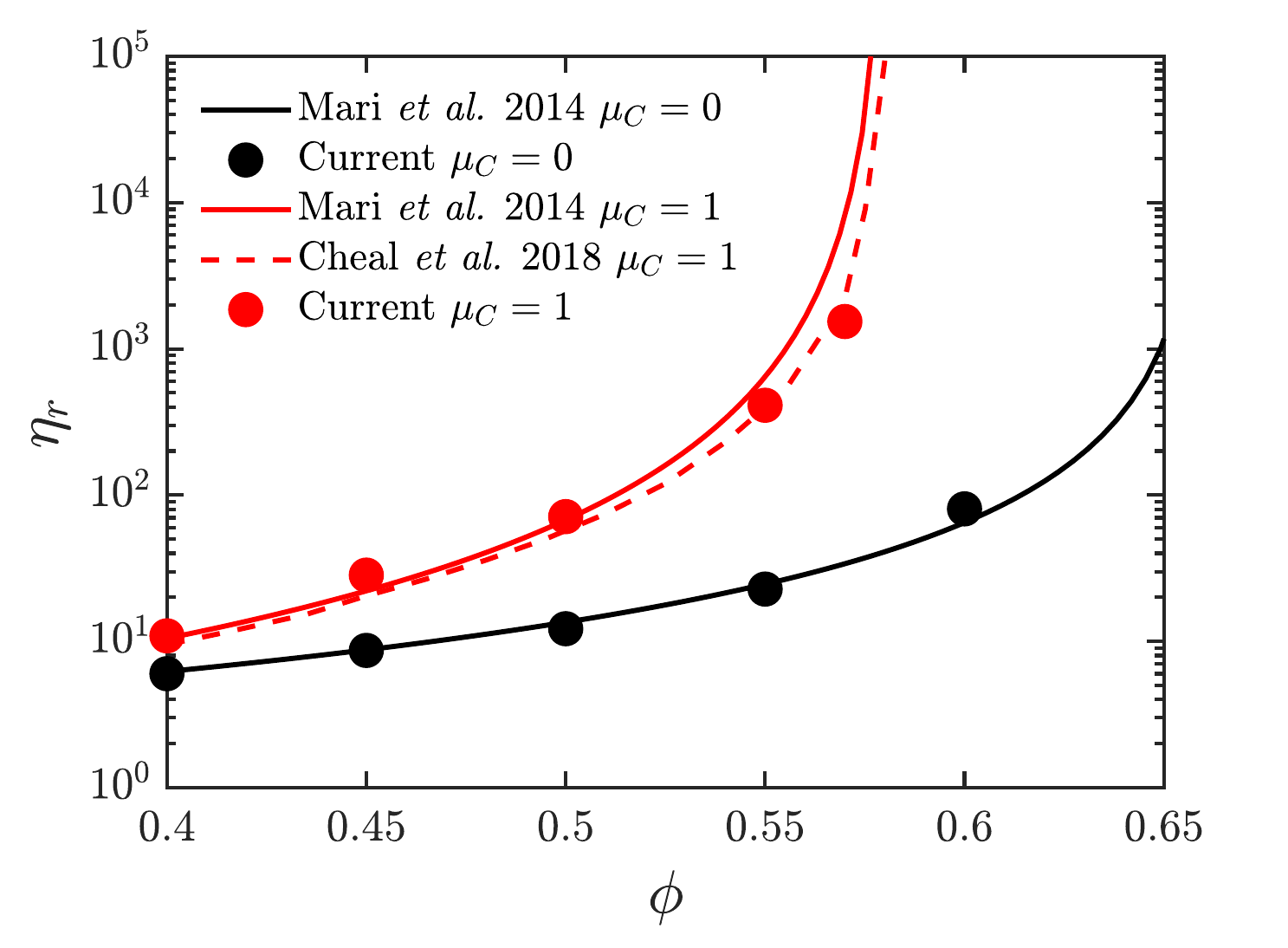}
    \caption{Relative viscosity, $\eta_r$ (filled markers), 
    varying the volume fraction. The solid curves represents 
    the power-law fitting of the data from \citep{MARI2014}. The black solid
    curve, $\eta_r=1.4\left( 1-\phi/\phi_J \right)^{-1.6}$, 
    with jamming volume fraction $\phi_J=0.66$, is obtained for frictionless
    particles, while the red solid curve, $\eta_r=0.71\left( 1-\phi / 
    \phi_J \right)^{-2.3}$, with jamming volume fraction $\phi_J=0.575$, 
    is obtained for frictional particles with $\mu_C = 1$. 
    The red dashed line, instead, is obtained from the numerical 
    data by Cheal and Ness \citep{CHEAL2018} for frictional particles 
    with $\mu_C=1$.}
    \label{fig:val}
\end{figure}

For the jamming transition, we consider a cubic box containing a 
bidispersed suspension ($a_2/a_1=1.4$, in equal volumes $V_2/V_1=1$) of $200$ 
frictionless, rigid (see \cref{eq:stiff}) particles, with no 
electro-chemical interactions. The side length of the box accommodates 
the volume fraction $\phi$ imposed, being $a_0=a_1=1$.
The parameters chosen echo the validation carried out by Ge and Brandt 
\citep{GE2020} and are listed in \cref{tab:param}.
\begin{table}
  \begin{center}
     \scalebox{0.8}{
     \setlength{\tabcolsep}{0.55em}
     \begin{tabular}{l|c|c|c|c|c|c|c|c|c|c}
         \hline
         \rule{0pt}{3ex}
         & $N$ & $a_2/a_1$ & $V_2/V_1$ 
         & $St$ & $\hat{k}$ & $k_t/k_n$ & $\gamma_n \dot{\gamma}/k_n$ 
         & $\mu_C$ & $\dot{\gamma}/\dot{\gamma}_0$ & $\delta_{lub}/a_0$ \\[2mm]
         FL & $200$ & $1.4$ & $1$ & $10^{-3}$ & $4\times 10^5$ & $2/7$ & $10^{-7}$
         & $0$ & $\infty$ & $\{0.05,0.2\}$ \\[2mm]
         FR & $1000$ & $1.4$ & $1$ & $10^{-3}$ & $4\times 10^5$ & $2/7$ & $10^{-7}$
         & $1$ & $\infty$ & $\{0.05,0.2\}$ \\ \hline
     \end{tabular}}
     \caption{Parameter choice for the frictionless (FL) and frictional (FR)
              configurations. Here, $a_0 = a_1$ represents the reference 
              length scale.}
     \label{tab:param}
   \end{center}
\end{table}
\Cref{fig:val} shows the comparison between the simulations results 
and the data available in the literature with a satisfying agreement between
the computed relative viscosity, $\eta_r$,  and the power-law fitting of 
Mari \textit{et al.} \citep{MARI2014} for frictionless particles 
(solid black curve).

We then activate the friction ($\mu_C = 1$) and we increase the 
number of particles to $N=1000$ (see \cref{tab:param}). \Cref{fig:val} 
shows the comparison between the computed relative viscosity $\eta_r$ 
and the data fitting obtained by Mari \textit{et al.} \citep{MARI2014} 
(solid red curve) and Cheal and Ness \citep{CHEAL2018} (dashed red curve) 
with excellent agreement.

The shear-dependent behaviour of the particles suspended in a uniform 
shear-flow is obtained by considering all contributions listed in 
\cref{sec:numMet}.
At first, we will show the results obtained from a bidispersed dense 
suspensions (size ratio $a_2/a_1=1.4$)
with volume fraction $\phi=[0.45,0.50,0.55]$, to see if the software is
able to catch the typical shear-dependent behaviour \citep{MARI2014}.
The set of parameters for carrying out these simulations have been chosen
to mimic those by \citet{MARI2014} and are listed in
\cref{tab:sr}.
\begin{table}
  \begin{center}
     \scalebox{0.8}{
     \setlength{\tabcolsep}{0.55em}
     \begin{tabular}{c|c|c|c|c|c|c|c|c|c|c|c|c}
         \hline
         \rule{0pt}{3ex}
         $\phi$ & $N$ & $a_2/a_1$ & $V_2/V_1$ 
         & $St$ & $\hat{k}$ & $k_t/k_n$ & $\gamma_n \dot{\gamma}/k_n$ 
         & $\mu_C$ & $\dot{\gamma}/\dot{\gamma}_0$ & $\delta_{lub}/a_0$
         & $\lambda_D/a_0$ & $F^R/F^A$ \\[2mm]
         $0.45$ & $512$ & $1.4$ & $1$ & $10^{-3}$ & $4\times 10^5$ & $2/7$ & $10^{-7}$
         & $1.0$ & $\{ 10^{-5},\dots,10\}$ & $0.10$ & $0.05$ & $\infty$ \\
         \rule{0pt}{3ex}
         $0.50$ & $512$ & $1.4$ & $1$ & $10^{-3}$ & $4\times 10^5$ & $2/7$ & $10^{-7}$
         & $1.0$ & $\{ 10^{-5},\dots,10\}$ & $0.10$ & $0.05$ & $\infty$ \\
         \rule{0pt}{3ex}
         $0.55$ & $512$ & $1.4$ & $1$ & $10^{-3}$ & $8\times 10^5$ & $2/7$ & $10^{-7}$
         & $1.0$ & $\{ 10^{-5},\dots,10\}$ & $0.10$ & $0.05$ & $\infty$ \\ \hline
     \end{tabular}}
      \caption{Set of parameters used for simulating the behaviour of
               dense suspensions varying the shear rate. 
               Here, $a_0 = a_1$ is the reference length scale.}
     \label{tab:sr}
   \end{center}
\end{table}
\begin{figure}[htb]
    \centering
    \includegraphics[width=0.32\textwidth]{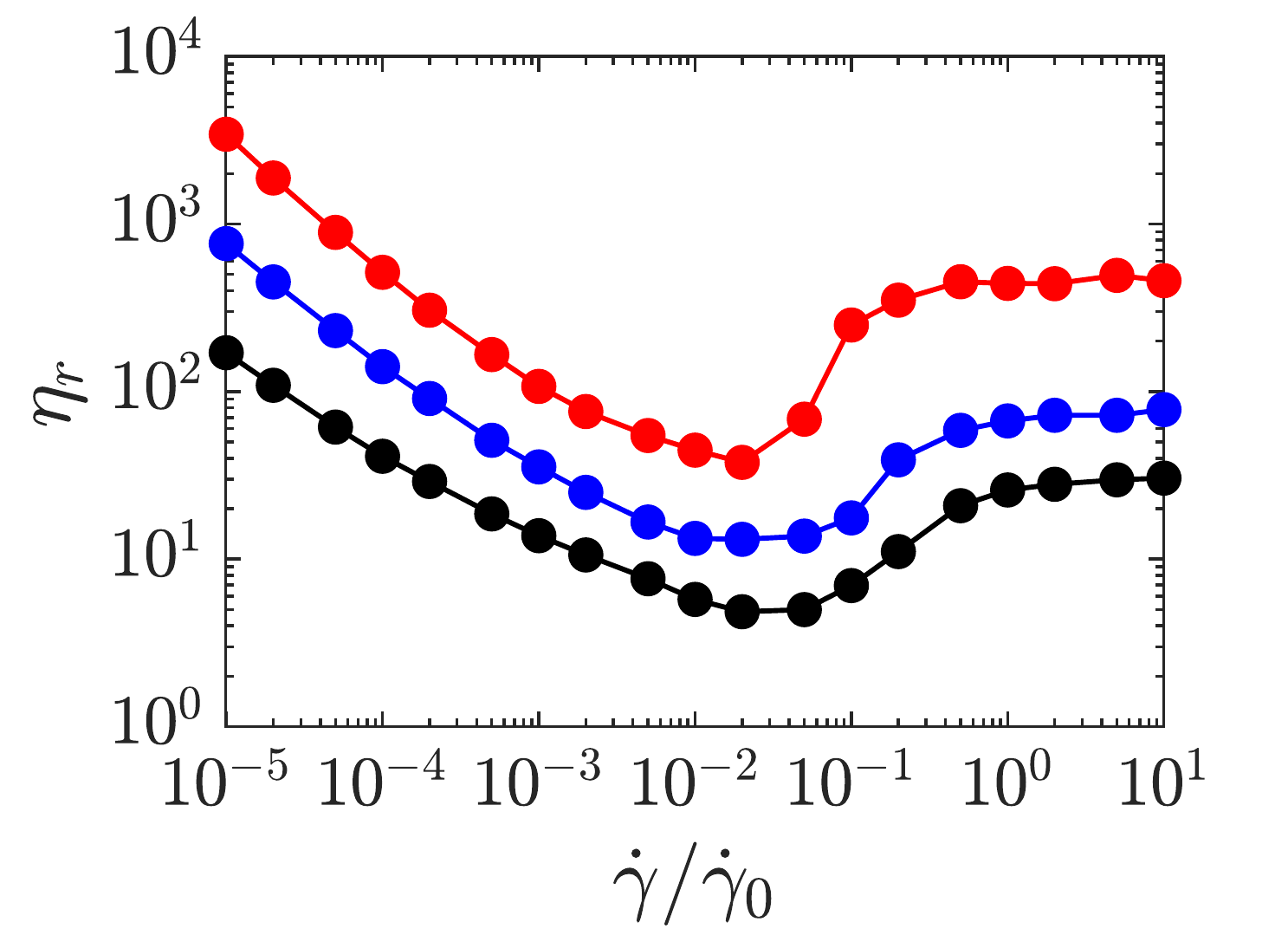}
    \includegraphics[width=0.32\textwidth]{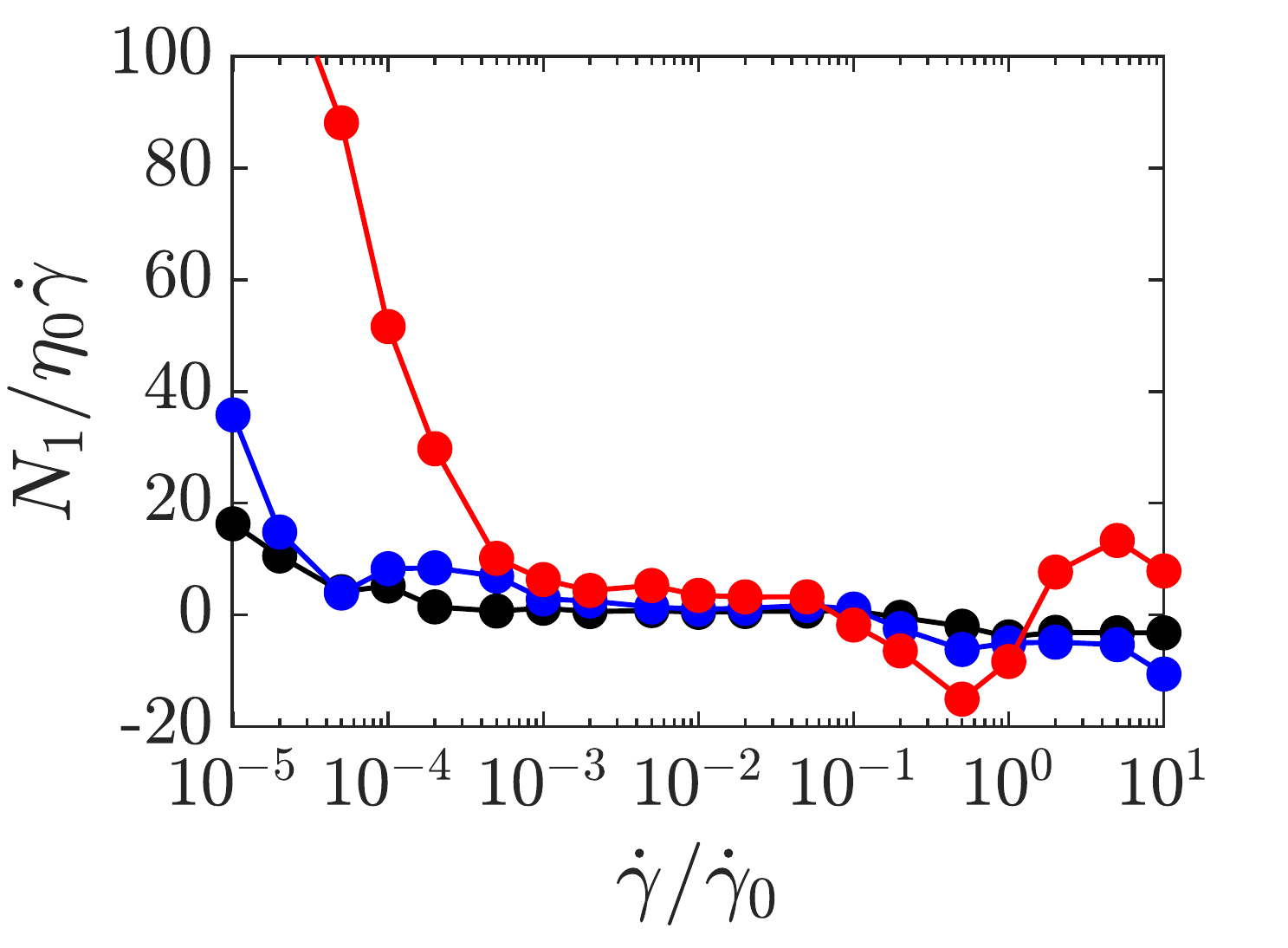}
    \includegraphics[width=0.32\textwidth]{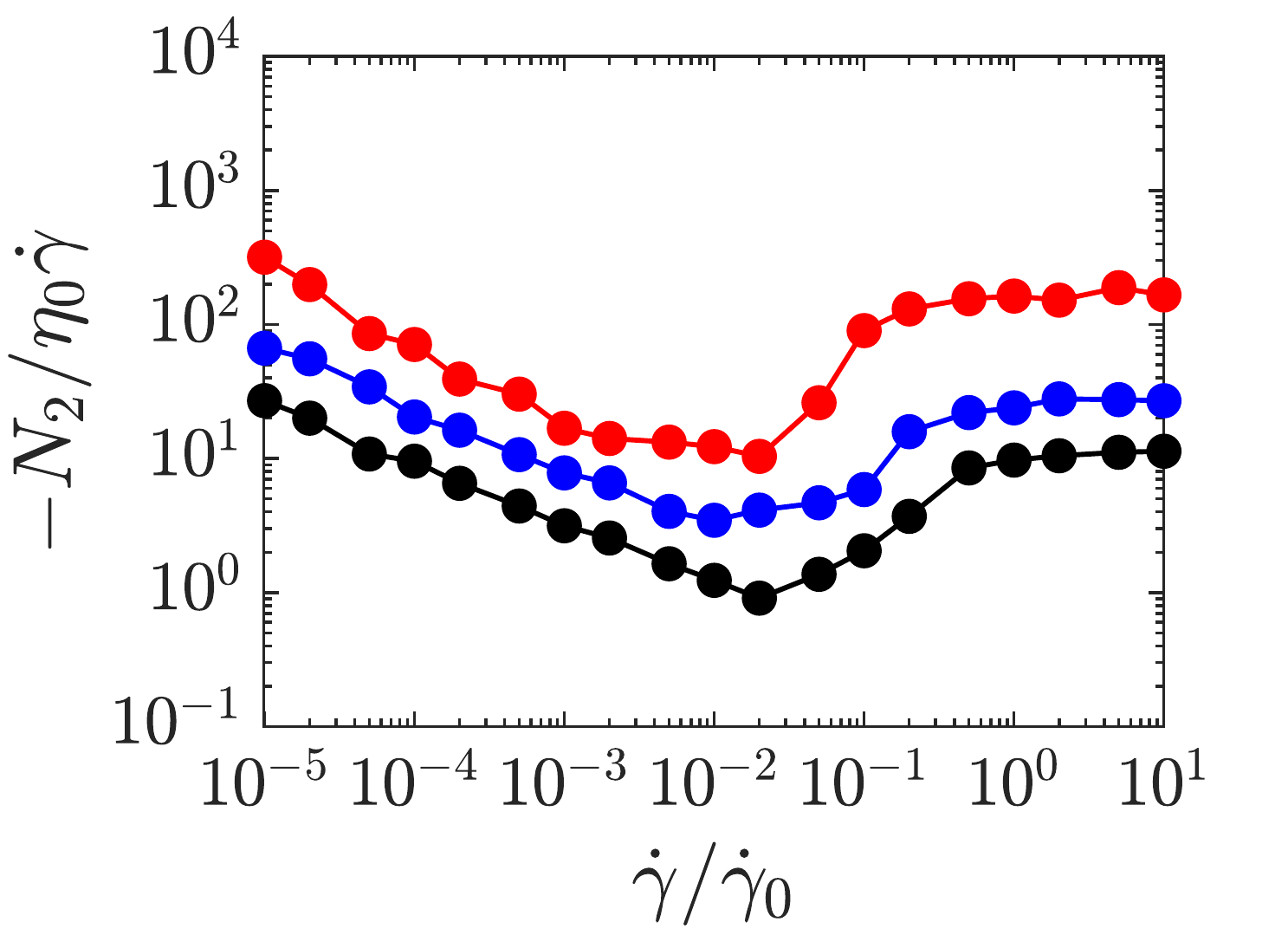}
    \caption{Shear-dependent rheological quantities of dense suspensions
    with $\phi=[0.45,0.50,0.55]$, indicated by the black, blue and red lines,
    respectively. Left panel: relative viscosity, $\eta_r$; 
    middle panel: first normal stress difference $N_1$
    normalized with the shear stress of the carrier fluid; right panel:
    Second normal stress difference $N_2$ normalized with the shear stress 
    of the carrier fluid.}
    \label{fig:etarSR}
\end{figure}
Note that for these simulations, the electro-chemical contribution is 
purely repulsive ($F_R/F_A=\infty$), as set in \citet{MARI2014}.
\Cref{fig:etarSR} (left panel) shows the trend of the relative viscosity $\eta_r$
varying the shear-rate applied to the suspension $\dot{\gamma}/\dot{\gamma}_0$.
In particular, we see that the three suspensions show the same qualitative behaviour, 
with values of the relative viscosity substantially increasing (especially at higher 
shear-rates) with the volume fraction.
From \cref{fig:etarSR}, we can notice that at low shear-rates the suspensions
show the typical shear-thinning behaviour with similar slope in the three cases considered,
as already observed by \citet{MARI2014}. At those shear-rates, the repulsive 
contribution dominates the mechanics of the suspensions (higher $F^E$, lower 
$\dot{\gamma}/\dot{\gamma}_0$), holding the particles off at a mid-range distance. 
On the other hand, at higher shear-rates the contacts become dominant, triggering the
typical shear-thickening behaviour, with a growing slope as the volume fraction
is increased, eventually leading to a quasi-discontinuous shear-thickening at 
$\phi=0.55$. When considering dense suspensions, to characterize the strain-induced 
anisotropy of the fluid, two additional rheologically meaningful functions are usually
measured: the two independent normal stress differences $N_1$ and $N_2$. 
\Cref{fig:etarSR} shows the first (middle panel) and the second (right panel) 
normal stress differences, respectively, of the shear-dependent suspensions 
considered (see \cref{tab:sr}).
In particular, the trend observed is similar to the one reported by \citet{MARI2014}, 
proving the capability of the software to catch the strain-induced anisotropies
of the suspensions.

\begin{figure}[htb]
    \centering
    \includegraphics[width=0.5\textwidth]{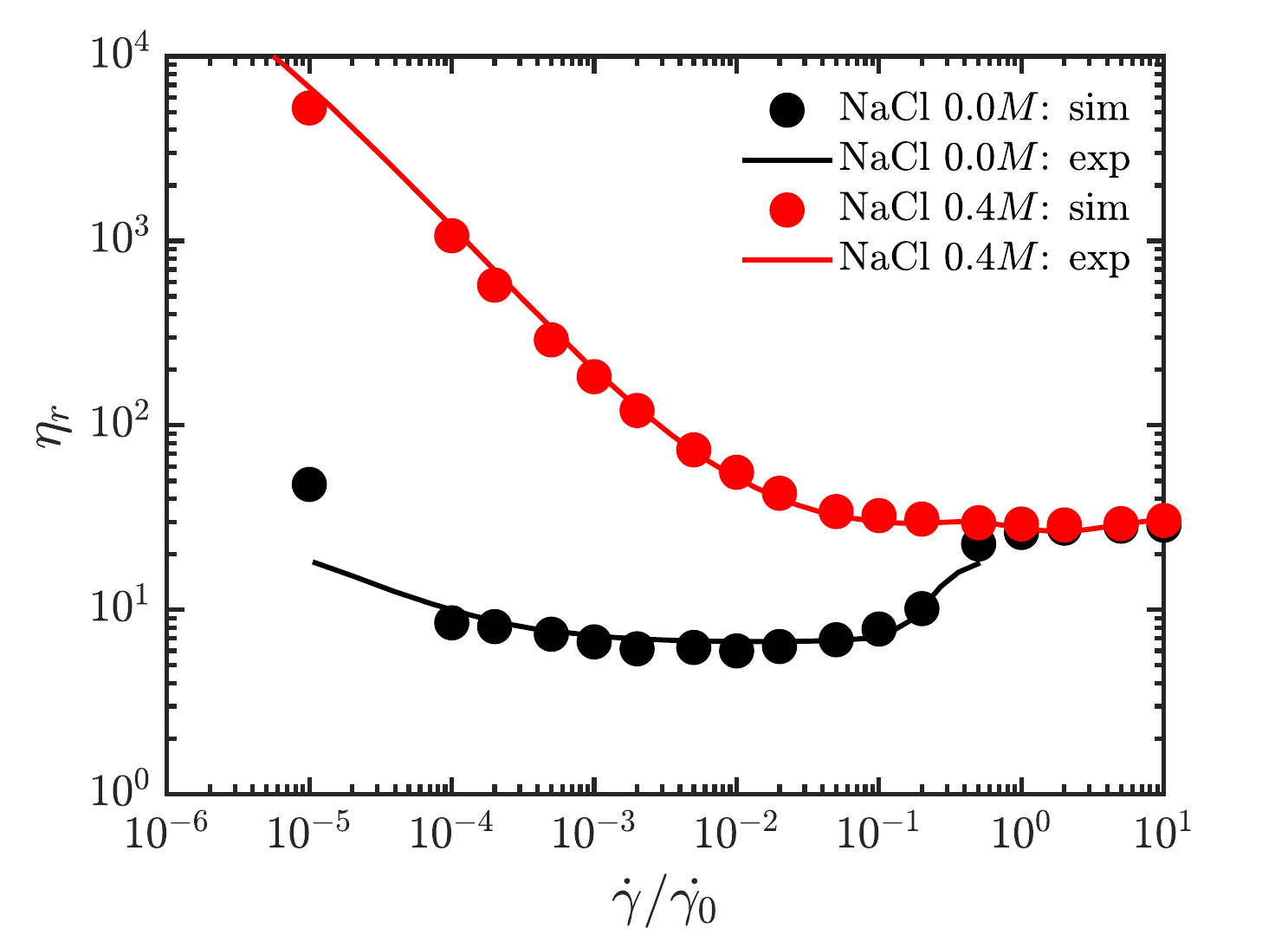}
    \caption{Relative viscosity, $\eta_r$, varying the shear-rate, $\dot{\gamma}/
    \dot{\gamma}_0$, where for the experiments, the value $\dot{\gamma}_0$ has been 
    chosen to match the shear-rate domain of the simulations. 
    The symbols outline the numerical data, while the solid lines represent 
    the experimental results. The black data are obtained from a salt-free 
    suspension, while the red data represent the solution
    with salt concentration $0.4M$.}
    \label{fig:valSR}
\end{figure}

Next, we consider a dense suspension with volume fraction $\phi=0.46$ and
we compare our numerical results with experimental data obtained from
quasi-neutrally buoyant silica colloidal particles, 
processed through the St{\"o}ber method \citep{STOEBER1968,ZHANG2009}, 
suspended in a glycerol-water solution with relative concentration 
$80\%-20\%$, respectively.
The electro-chemical properties of the suspension were altered
by adding sodium-chloride (NaCl) to the solution, as done in \citet{RATHEE2021}.
In particular
The numerical and experimental parameters are reported in \cref{tab:exp}.
\begin{table}
  \begin{center}
     \scalebox{0.8}{
     \setlength{\tabcolsep}{0.55em}
     \begin{tabular}{l|c|c|c|c|c|c|c|c|c|c|c|c}
         \hline
         \rule{0pt}{3ex}
         & $N$ & $a_2/a_1$ & $V_2/V_1$ 
         & $St$ & $\hat{k}$ & $k_t/k_n$ & $\gamma_n \dot{\gamma}/k_n$ 
         & $\mu_C$ & $\dot{\gamma}/\dot{\gamma}_0$ & $\delta_{lub}/a_0$
         & $\lambda_D/a_0$ & $F^R/F^A$ \\[2mm]
         SI & $512$ & $1.09$ & $1$ & $10^{-3}$ & $4\times 10^5$ & $2/7$ & $10^{-7}$
         & $0.5$ & $\{ 10^{-5},\dots,10\}$ & $0.25$ & $\left[0.0225,0.0065\right]$ & $\left[19,1\right]$ \\ \hline
         \rule{0pt}{3ex}
         & \multicolumn{2}{c|}{$a_0$}          & \multicolumn{2}{c|}{$\eta_0$} 
         & \multicolumn{2}{c|}{$\rho_{fluid}$} & \multicolumn{2}{c|}{$\mu_C$} 
         & \multicolumn{2}{c|}{$\lambda_D$}    & \multicolumn{2}{c}{$NaCl$} \\[2mm]
         EX 
         & \multicolumn{2}{c|}{$208\pm 5\,nm$} & \multicolumn{2}{c|}{$0.108\,Pa\,\,s$} 
         & \multicolumn{2}{c|}{$1.12\,g/cm^3$} & \multicolumn{2}{c|}{$0.5$} 
         & \multicolumn{2}{c|}{$\left[4.5,1\right]\,nm$}
         & \multicolumn{2}{c}{$\left[0.4,0\right]\,M$} \\ \hline
     \end{tabular}}
      \caption{Parameters set used for numerical simulations (SI) and 
              experiments (EX) of particles suspended in a varying
              uniform shear-flow. Here, $a_0$ ($= a_1$ in the simulations) 
              stands for the reference length scale.}
     \label{tab:exp}
   \end{center}
\end{table}
\Cref{fig:valSR} shows a very good agreement between the relative viscosity, 
$\eta_r$, obtained from the numerical and experimental data, for both salt
concentrations considered. For the experimental results, the shear-rate 
$\dot{\gamma}$ has been normalised to match the domain span of the numerical 
data.
The two suspensions behave very differently at lower shear-rates, i.e
in the shear-thinning region. In that regime, the electro-chemical
potential acting on the suspension dominates its mechanics.
Increasing the salt concentration, the electro-chemical properties of the suspension
are modified, with the contribution due to the van der Waals attractive forces becoming
more important (lower $F_R/F_A$). This induces the formation of large clusters that highly
contribute to the higher resistance of the suspension to flow \citep{RATHEE2021} and
this is reflected in \cref{fig:valSR}, with the red curve showing much higher values
of $\eta_r$.
Conversely, at higher shear-rates, where the contacts between particles dominate, 
the behaviour of the suspension does not depend on the salt concentration; therefore,
the two suspensions have similar values of relative viscosities (see \cref{fig:valSR})
\citep{RATHEE2021}.

\begin{figure}[htbp]
    \centering
    \includegraphics[width=0.49\textwidth]{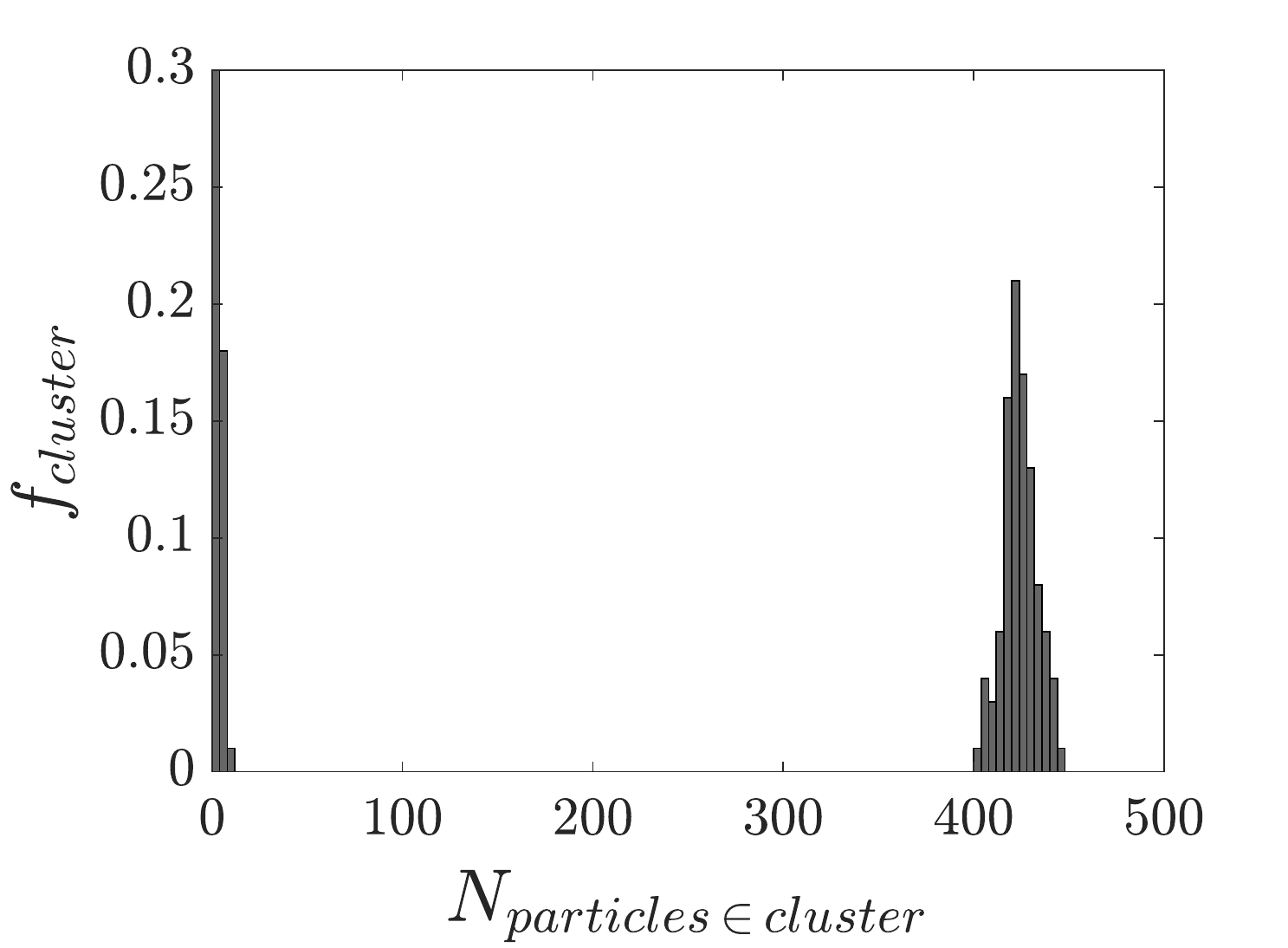}
    \includegraphics[width=0.49\textwidth]{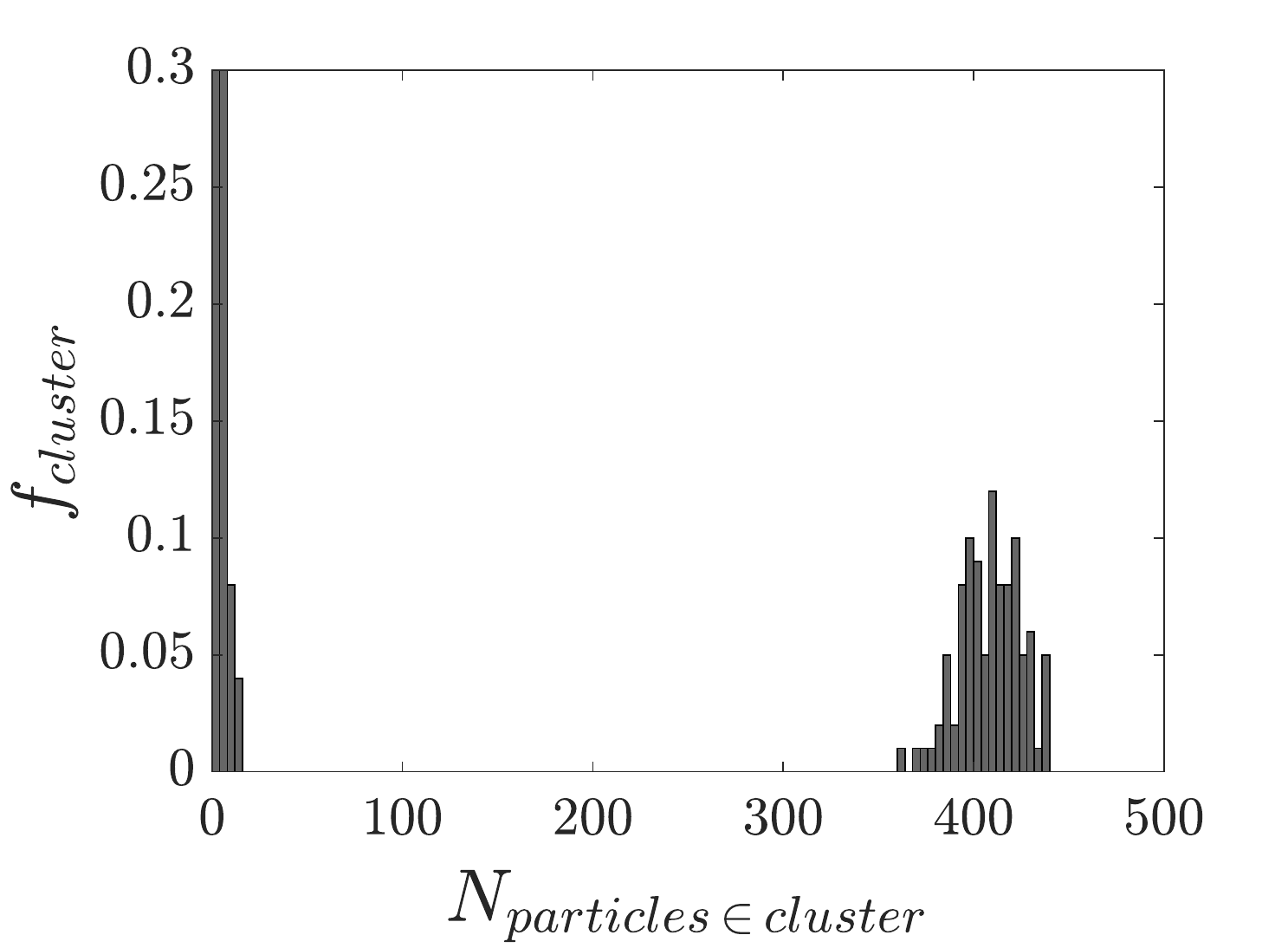}
    \caption{Average size distribution of clusters of contacting particles
             for the strongly attractive suspension (i.e. salt concentration
             $0.4M$). Left panel: $\dot{\gamma}/\dot{\gamma}_0=10^{-5}$; 
             right panel: $\dot{\gamma}/\dot{\gamma}_0=10^0$.
             The total number of particles is $N=512$.
             The size distributions are averaged over $100$ snapshots.}
    \label{fig:histo}
\end{figure}

\begin{figure}[htbp]
    \centering
    \includegraphics[width=0.9\textwidth]{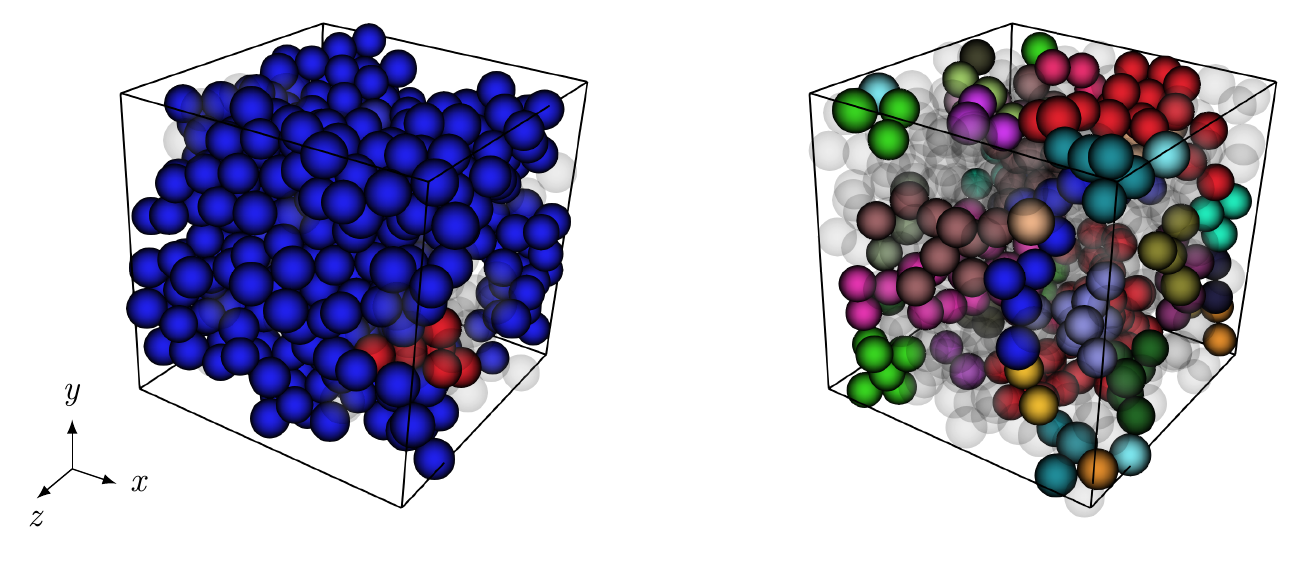}
    \caption{Snapshot of the cluster of contacting particles at 
             $\dot{\gamma}/\dot{\gamma}_0=10^{-5}$.
             Left panel: steady-state regime; right panel: transient regime.
             The total number of particles is $N=512$. The smaller cluster 
             plotted has a size of $5$ particles; smaller clusters are
             not shown for clarity of the images.}
    \label{fig:cluster}
\end{figure}

The typical size of the cluster of particles for the suspension with higher salt 
concentration is shown in \cref{fig:histo}. In particular, the two panels show the
size distribution of the clusters at the extrema of the curve in \cref{fig:valSR}, 
i.e. at shear-rate $\dot{\gamma}/\dot{\gamma}_0=[10^{-5},10^0]$, when the suspension
reaches its steady-state regime.
From \cref{fig:histo}, we can see that the distributions show two peaks,
the leftmost one concerning small clusters and the rightmost large ones.
Comparing the distributions of the latter at the two shear-rates considered,
we can see that the peak is narrower in the region where the van der Waals
forces dominate the behaviour of the suspension (left panel), with the particles 
gathering in a spherical-shaped cluster. A snapshot of the structures for the
case with salt concentration $0.4M$ is shown in \cref{fig:cluster} (left panel), 
where the two peaks seen in the distributions can be clearly recognised.
In the right panel of \cref{fig:cluster}, instead, a snapshot during the time 
evolution of the clusters is shown, with the large cluster fragmented
in many smaller ones.

\subsection{Wavy shear-flow}
The examples discussed above show that the code is able to 
tackle particles suspended in a uniform shear-flow. We extend here
the simulations to non-uniform shear-flows that activate the Fax{\'e}n
contribution introduced in the Stokes drag. In particular, we
introduce a wavy shear-flow,
\begin{equation}
    \protect{\boldsymbol{U}^{\infty} =}
    \begin{pmatrix}
        \dot{\gamma}y + \mathcal{U} \sin{\left(\dfrac{2\pi n}{L_y} y\right)} \\
        0 \\
        0
    \end{pmatrix},
    \label{eq:uinf}
\end{equation}
where $\mathcal{U}$ is the amplitude of the wave and $n$ 
imposes the number of spatial periods within the domain. 
This is a condition of relevant interest as it can be used 
for example to reproduce the wall effects that appear in a 
rheometer when attempting to replicate a uniform shear-flows 
or to take into account the back-reaction of the particles 
on the flow itself.

\begin{figure}[htbp]
    \centering
    \includegraphics[width=0.5\textwidth]{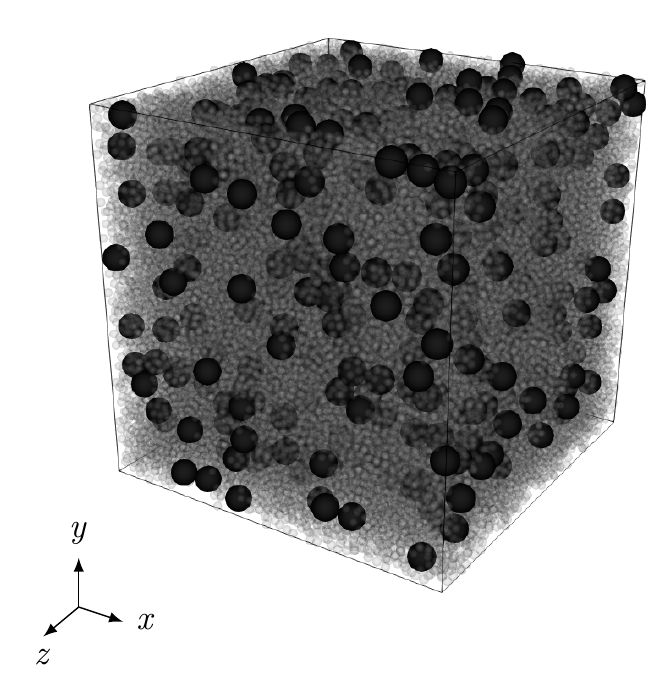}
    \caption{Snapshot of the large domain simulation with $N=65536$ and
             $a_2/a_1=3$. The volume fraction is $\phi=0.5$, with a volume
             contribution of the bidispersed particles $V_2/V_1=0.25$.}
    \label{fig:largeN}
\end{figure}

We consider a cubic box containing a bidispersed suspension with volume 
fraction $\phi=0.5$. The ratio between the radii of the two dispersed 
phases is $a_2/a_1 = 3.0$, while the volume percentage ratio of the largest 
dispersed particles to the smallest one is $V_2/V_1 = 0.25$. 
We set the number of particles to $N=65536$ to test the memory-efficiency 
of the software.  The side length of the box accommodates the volume fraction 
$\phi=0.5$, being $a_0=a_1=1$.
\begin{table}[htbp]
  \begin{center}
     \scalebox{0.8}{
     \setlength{\tabcolsep}{0.55em}
     \begin{tabular}{l|c|c|c|c|c|c|c|c|c|c|c|c|c|c}
         \hline
         \rule{0pt}{3ex}
         & $N$ & $a_2/a_1$ & $V_2/V_1$ 
         & $St$ & $\hat{k}$ & $k_t/k_n$ & $\gamma_n \dot{\gamma}/k_n$ 
         & $\mu_C$ & $\dot{\gamma}/\dot{\gamma}_0$ & $\delta_{lub}/a_0$ 
         & $\lambda_D/a_0$ & $F^R/F^A$ & $\mathcal{U}/\dot{\gamma}a_0$ & $n$ \\[2mm]
         NW & $2^{16}$ & $3$ & $0.25$ & $10^{-3}$ & $4\times 10^5$ & $2/7$ & $10^{-7}$
         & $1$ & $1$ & $0.25$ & $0.02$ & $19$ & $0$ & $0$ \\[2mm]
         YW & $2^{16}$ & $3$ & $0.25$ & $10^{-3}$ & $4\times 10^5$ & $2/7$ & $10^{-7}$
         & $1$ & $1$ & $0.25$ & $0.02$ & $19$ & $20$ & $1$ \\ \hline
     \end{tabular}}
     \caption{Parameters set for dense suspensions in a uniform shear-flow
              (NW) and in a wavy one (YW), following \cref{eq:uinf}.
              Here, $a_0 = a_1$ represents the reference length scale.}
     \label{tab:wavy}
   \end{center}
\end{table}
The particles composing the suspension are considered to be rigid with
a non-smooth surface (i.e. with a friction coefficient $\mu_C = 1$), electrically
charged with a screening length $\lambda_D=0.02 a_0$. The equivalent shear-rate,
\cref{eq:ec}, is set to $\hat{\dot{\gamma}} = 1$ (i.e. within the 
shear-thickening region of \cref{fig:valSR}), with a ratio $F^R/F^A=19$ when 
the pair of particles $(i,j)$ are at contact.
The wave was chosen to have a single spatial period, $n=1$, with a large 
amplitude $\mathcal{U} = 20 \dot{\gamma} a_0$ to amplify the effects of 
the wavy shear.
\Cref{tab:wavy} wraps up the set of parameters chosen and a snapshot of 
the domain and particles is shown in \cref{fig:largeN}.

\begin{figure}[htbp]
    \centering
    \includegraphics[width=0.49\textwidth]{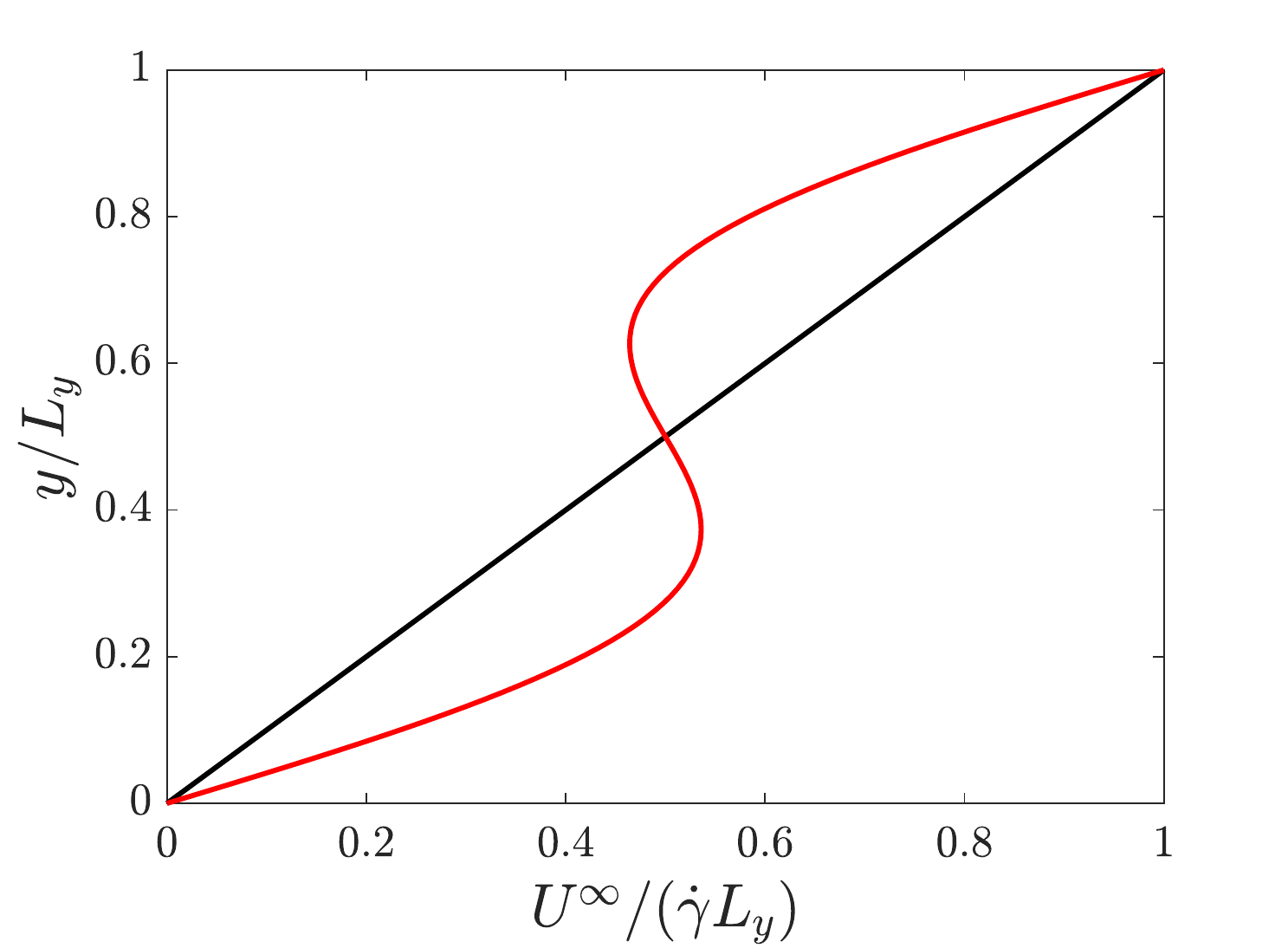}
    \includegraphics[width=0.49\textwidth]{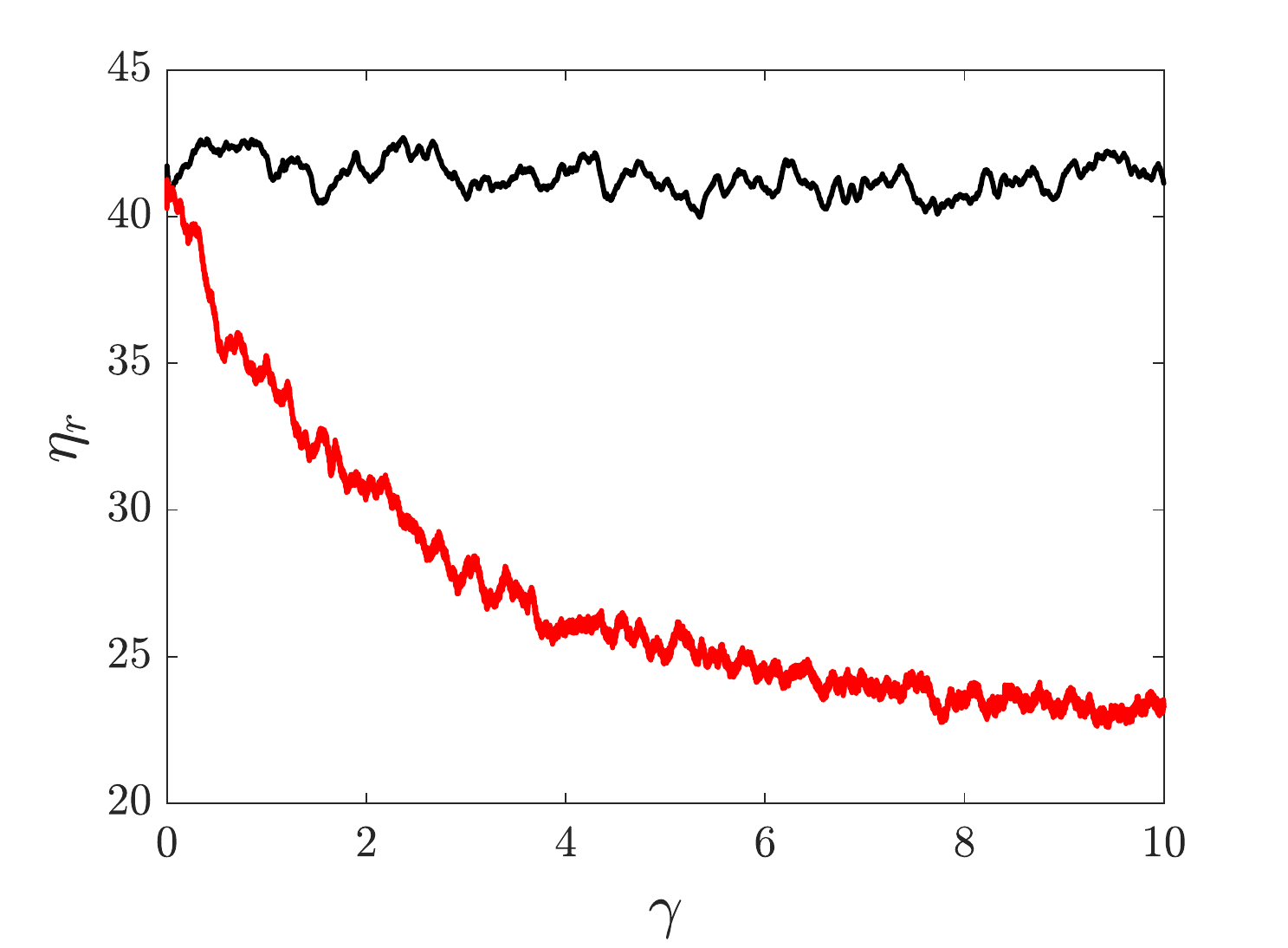}
    \caption{Left panel: streamwise velocity profile of the carrier flow
             applied in the shear-wise direction, $y$. Right panel:
             time history of the relative viscosity $\eta_r$, being 
             $\gamma=t \dot{\gamma}$ the number of strains.
             In both panels, the black line represents the suspension
             in a uniform shear-flow, while the red line the suspension
             in a wavy one.}
    \label{fig:timehis}
\end{figure}

A wave with a large amplitude, such as the one applied in the case described above,
significantly affects the local shear-rate, as shown in the left panel of 
\cref{fig:timehis}.
This local effect has a direct impact on the suspension as half of it is 
located in a low shear-rate region, while the other half feels a high 
shear-rate. Thus, the non-uniformity impacts the rheology of the suspension 
modifying the relative viscosity $\eta_r$ of the suspension, as shown 
in the right of \cref{fig:timehis}. Note that both cases analysed
started from an identical initial condition.

\begin{figure}[htbp]
    \centering
    \includegraphics[width=0.49\textwidth]{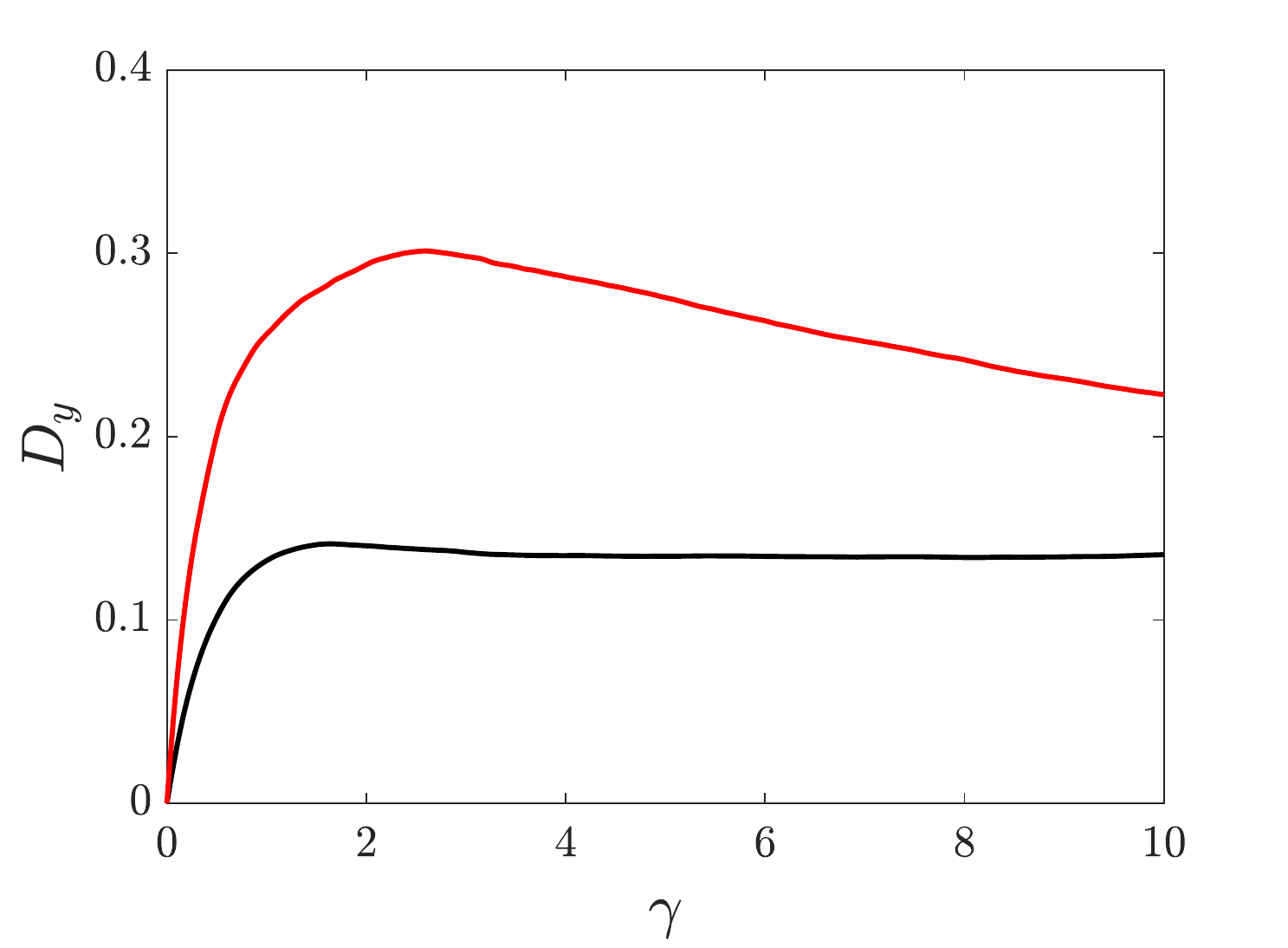}
    \includegraphics[width=0.49\textwidth]{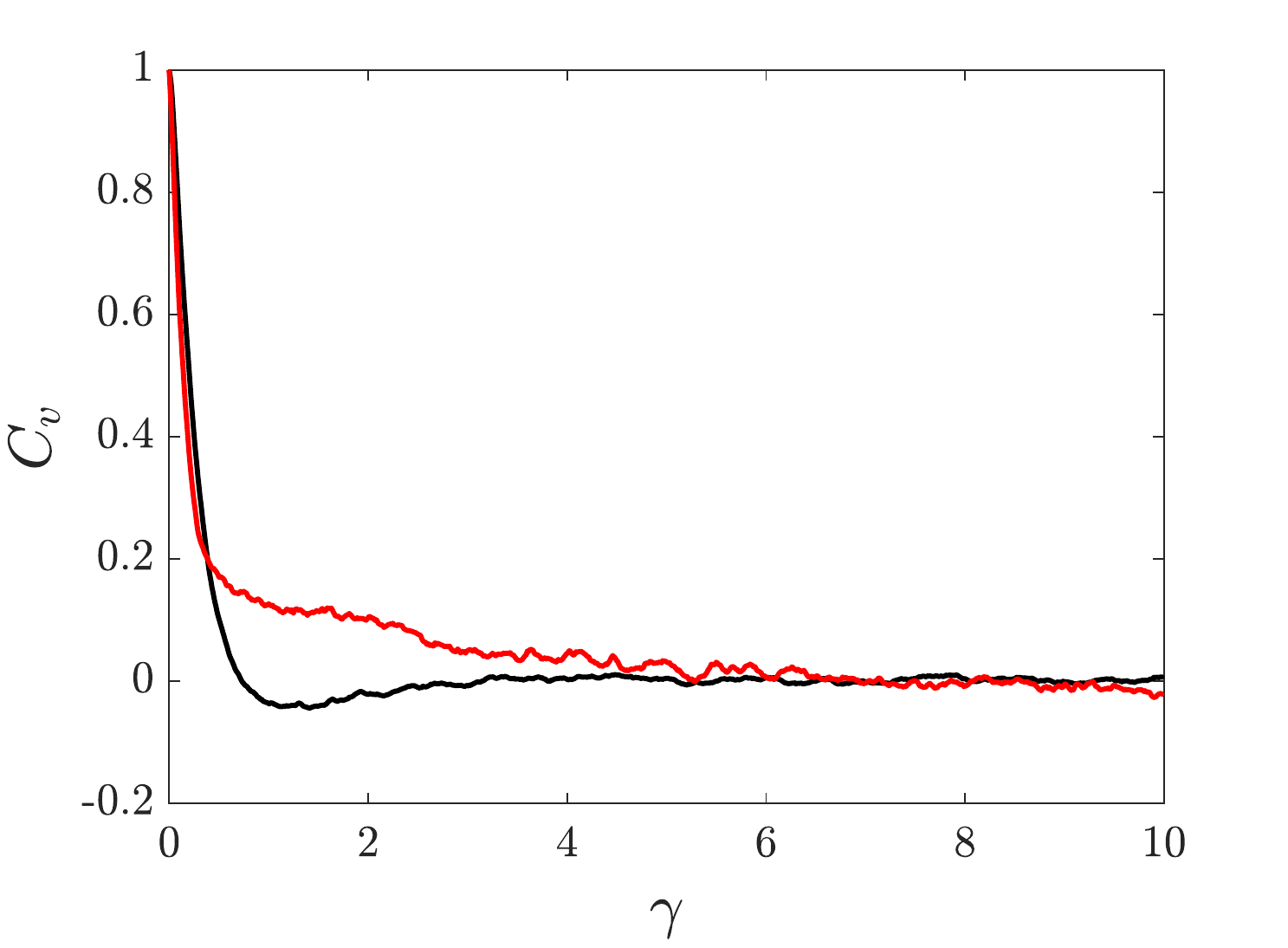}
    \caption{Left panel: time-history of the shear-wise ($y$-direction)
             diffusivity of the particles. Right panel:
             time history of the autocorrelation function of the shear-wise
             velocity fluctuations $v'$. The non-dimensional time is plotted 
             in term of strains, $\gamma=t \dot{\gamma}$.
             In both panels, the black line represents the suspension
             in a uniform shear-flow, while the red line the suspension
             in a wavy one.}
    \label{fig:diff}
\end{figure}

Next, in \cref{fig:diff} we plot the shear-wise diffusivity (left panel) and
the autocorrelation function of the shear-wise velocity fluctuations (right panel), 
respectively defined as
\begin{equation}
    D_y = \dfrac{1}{a_0^2\dot{\gamma}}\dfrac{\langle \left[y(t)-y(t_0)\right]^2\rangle}{2(t-t_0)},
    \label{eq:diffY}
\end{equation}
and
\begin{equation}
    C_v = \dfrac{\langle v'(t) v'(t_0)\rangle}{\langle v'(t)^2\rangle},
    \label{eq:corrV}
\end{equation}
where in both equations, the $\langle\,\rangle$ brackets indicate the average 
over all the particles and $t_0$ is the initial time; $y(t)$ and $v'(t)$ are
the time-dependent shear-wise location and velocity fluctuations, respectively,
of the particles. The figure shows how the particles, on average, diffuse more 
(twice more) when the wavy-shear rate is applied to the suspension, due to the 
non-uniformity of the shear rate. Moreover, the autocorrelation function of $v'$ 
shows a higher integral time-scale for the suspension with the non-uniform
shear-rate.

\begin{figure}[htbp]
    \centering
    \includegraphics[width=0.32\textwidth]{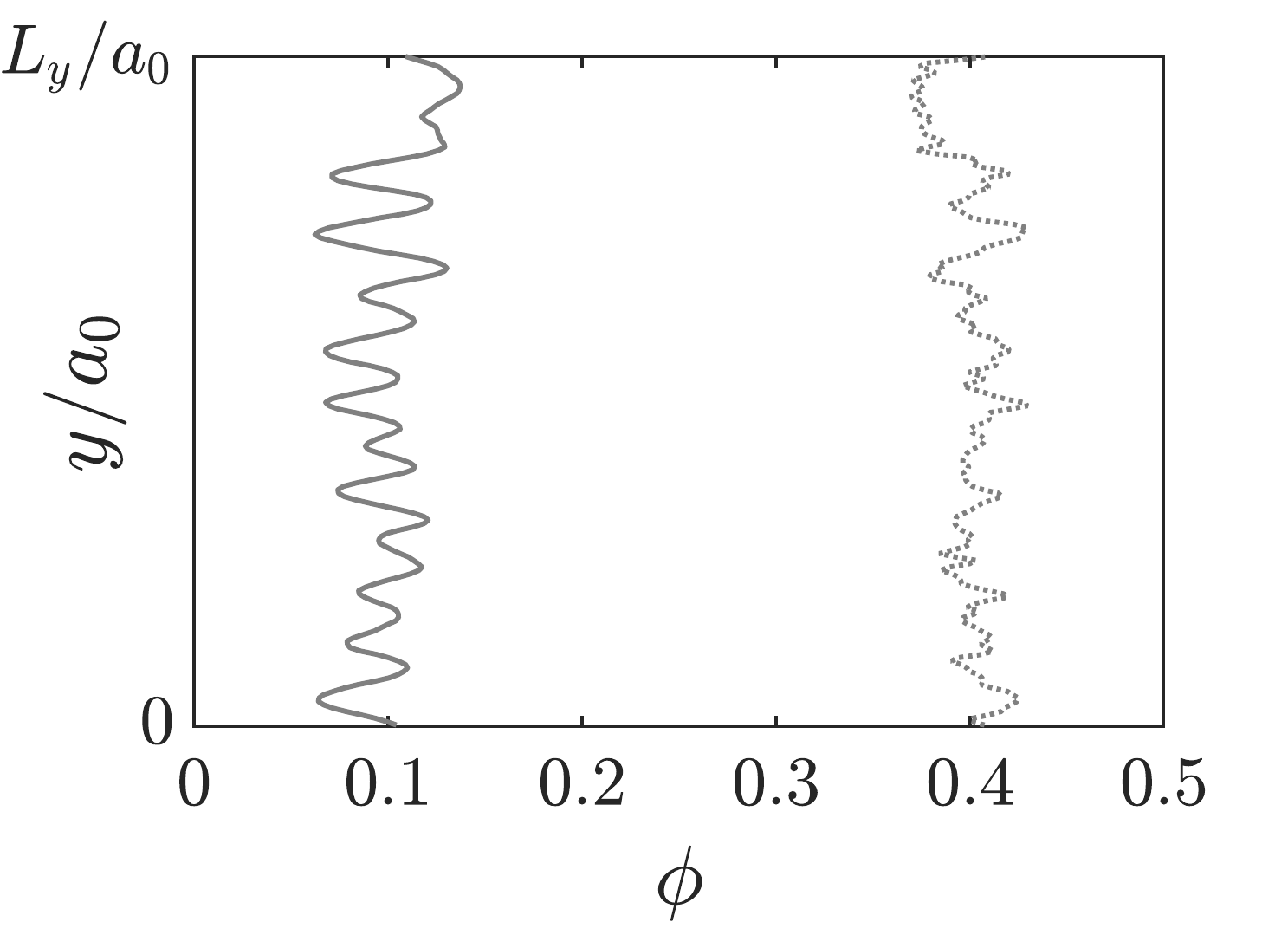}
    \includegraphics[width=0.32\textwidth]{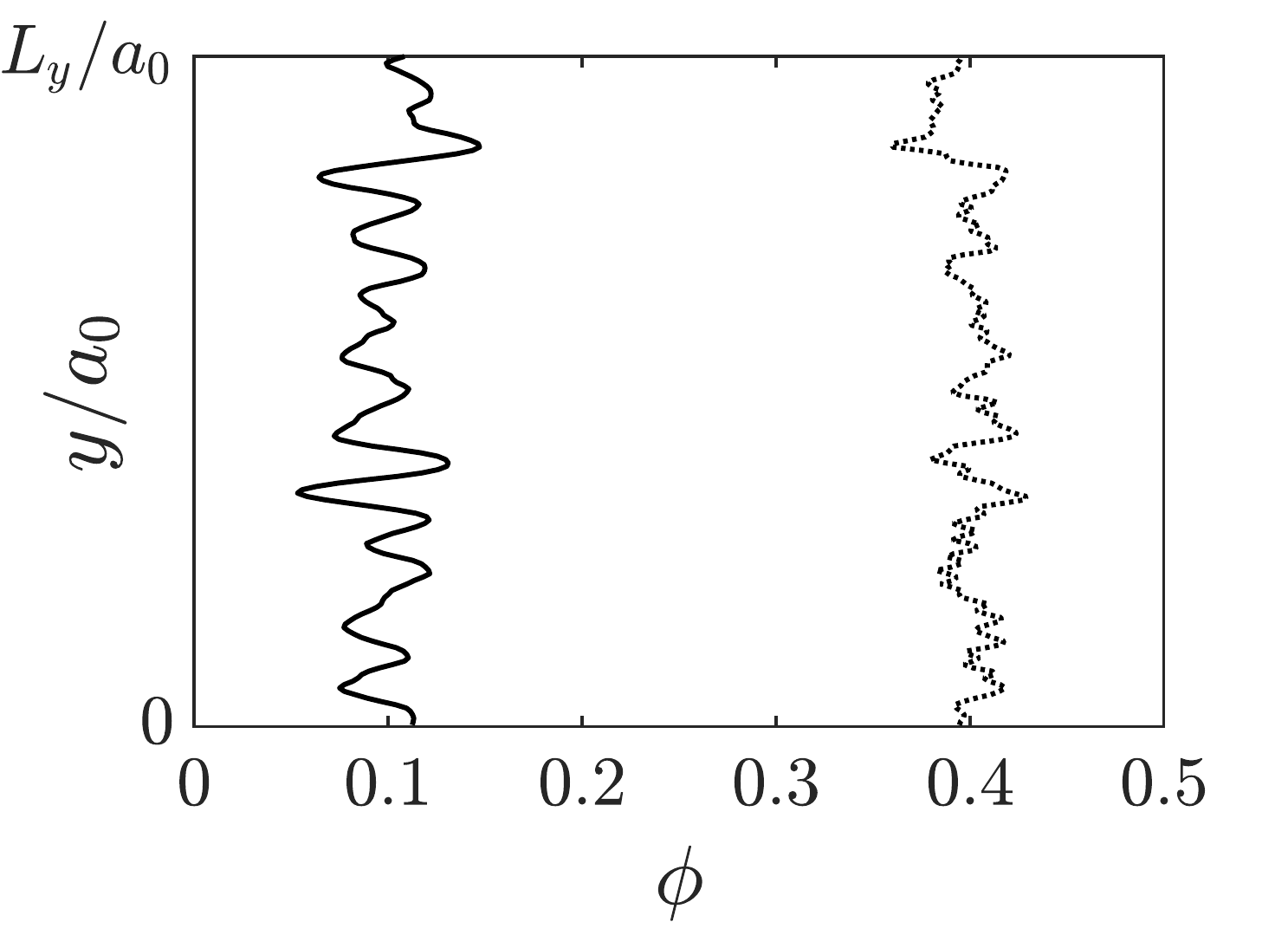}
    \includegraphics[width=0.32\textwidth]{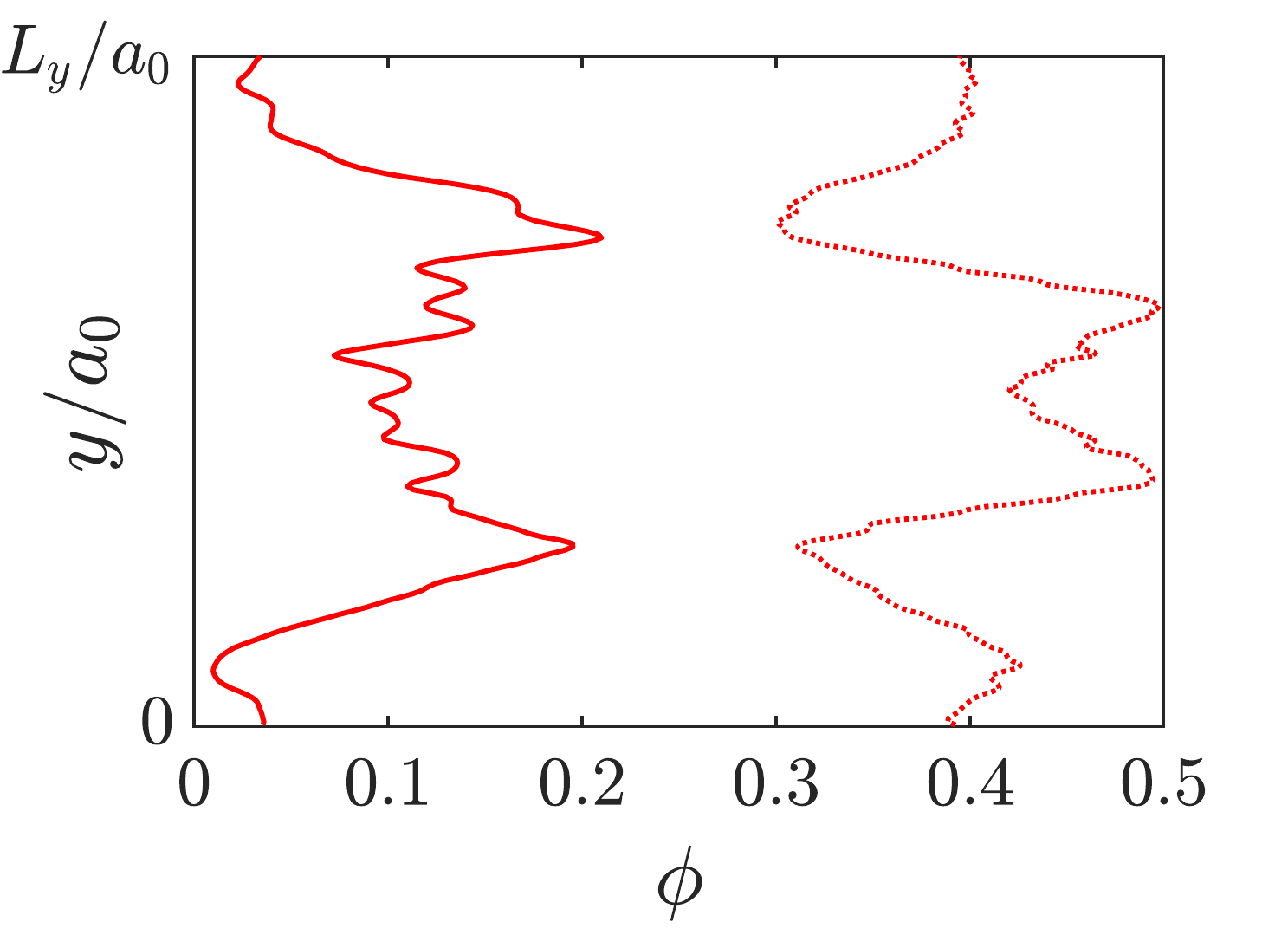}
    \caption{Local distribution of the volume fraction along the
             shear-wise direction. The solid lines represent the
             distribution of the large particles, while the dotted lines
             the distribution of the small particles.
             Left panel: initial condition; middle panel: uniform
             shear-rate; right panel: wavy shear-rate.
             The distributions are averaged over $100$ snapshots.}
    \label{fig:localPhi}
\end{figure}

Finally, to understand the temporal behaviour of the suspension, we show the
distribution of the local volume fraction along the shear-wise direction.
\Cref{fig:localPhi} shows the distribution of the particles in the shear-wise
direction separated by the size of the suspended phase. 
In particular the figure is organised as
follows: in the left panel, the local distributions of the large (solid line) 
and the small (dotted line) particles of the suspension at the initial time
are shown; in the middle panel, the same quantities are shown for the particles
suspended in a uniform shear-flow at $\gamma=10$; in the right panel, the local
distributions are shown for the particles suspended in a wavy shear-flow at
$\gamma=10$. While the distribution of the particles in the case where a 
uniform shear-rate is applied to the suspension remains similar to the initial
condition with the particles uniformly distributed along the
shear-wise direction, in the case where the particles are suspended in
a non-uniform shear flow, two clear peaks in the distribution of the large 
particles appear. This suggests that the large particles are migrating toward
such region, that in particular coincide with the location of zero-shear.

\section{Summary and outlook}
\label{sec:concl}
We have presented an efficient software 
(freely available at \url{https://github.com/marco-rosti/CFF-Ball-0x}) 
that tackles the dynamics of dense suspensions. First, the models of the 
contributions adopted to simulate the particle-particle and the particle-flow 
interactions have been presented. The models mimic the ones already available
in the literature, echoing in particular the implementation of 
Mari \textit{et al.} \citep{MARI2014} and Ge \textit{et al.} \citep{GE2020,GE2020b}. 
The core of the algorithm, i.e. the fixed-radius near neighbours search, 
follows the idea of Hoetzlein \citep{HOETZLEIN2014} 
with an improved symmetric search that guarantees a speed-up in the near 
neighbours inquiry and a good load-balance among processes when the 
parallelisation is adopted. Particular attention has been also given to 
the memory efficiency aspect of the implementation as the representation 
of rheological suspension with a large number of non-smooth particles is 
of high interest, especially to cope with particle suspensions having high size 
ratio, as a limited number of large particles would fit into small domains.

The aforementioned features of the implementation have been proved in the
examples carried out in \cref{sec:res}. First, we verified the 
reliability of the method by comparing our results with data available
in the literature obtained from other codes or experiments. Then, we carried out
sample simulations with a high number of particles $N=2^{16}$, comparing the
rheology of suspensions driven by a uniform and a wavy shear-flow.

The present code can be extended in a series of ways. Models in consideration 
for future implementations concern the possibility of reproducing Brownian 
particles and the introduction of walls that constrains the particles 
in any of the three directions. The present code is meant to be used
for studying the rheology of particle suspensions in several configurations, 
to unravel some of the complex phenomena arising when dealing 
with dense suspensions \citep{MORRIS2020}.

\section*{Acknowledgement}
All authors gratefully acknowledge the support of Okinawa
Institute of Science and Technology Graduate University (OIST)
with subsidy funding from the Cabinet Office, Government of
Japan. The authors also acknowledge the computer time provided 
by the Scientific Computing section of Research Support Division at OIST. 
A. Q. S. acknowledges funding from
the Joint Research Projects (JRPs) supported by JSPS and SNSF.
V. R. also acknowledges financial support from the Japanese
Society for the Promotion of Science (Grants No. 21K13895).

\section*{Conflicts of interest}
The authors have no conflicts to disclose.

\appendix
\section{}\label{app:lub}
The hydrodynamics of a dense suspension is dominated by contribution of
the close approaching particles \citep{BALL1997}. In a Stokesian framework, 
this translates in dropping the full description of the flow and modelling the
hydrodynamics contribution acting on the suspension with only close-range terms, 
described by the resistances matrices in \cref{eq:hydro}.

Following the description in Mari \textit{et al.} \citep{MARI2014} and 
adding the Fax{\'e}n laws \citep{GUAZZELLI2011}, the matrices for rigid 
spheres can be written as
\begin{equation}
    \label{eq:rst}
    \mathbb{R}_{Stokes,ij} =
    \begin{pmatrix}
      6\pi \eta_0 a_i (1+a_i^2/6\nabla^2) & 0 & 0 & 0   \\
      0 & 6\pi \eta_0 a_j (1+a_j^2/6\nabla^2) & 0 & 0   \\
      0 & 0 & 8\pi \eta_0 a_i^3 & 0 \\
      0 & 0 & 0 & 8\pi \eta_0 a_j^3
    \end{pmatrix},
\end{equation}

\begin{equation}
    \label{eq:rlu}
    \mathbb{R}_{Lub,ij} =
    \begin{pmatrix}
        X^A_{ii} \mathbb{N} + Y^A_{ii} \mathbb{T} & 
        X^A_{ij} \mathbb{N} + Y^A_{ij} \mathbb{T} &
        Y^B_{ii} \mathbb{N}_{\times} & Y^B_{ji} \mathbb{N}_{\times}\\
        & 
        X^A_{jj} \mathbb{N} + Y^A_{jj} \mathbb{T} &
        Y^B_{ij} \mathbb{N}_{\times} & Y^B_{jj} \mathbb{N}_{\times}\\
        \text{sym} & &
        Y^C_{ii} \mathbb{N}_{\times} & Y^C_{ij} \mathbb{N}_{\times}\\
        & & & Y^C_{jj} \mathbb{N}_{\times}\\
    \end{pmatrix},
\end{equation}

\begin{equation}
    \label{eq:rlu1}
    \mathbb{R}_{Lub,ij}^{'\left(\mathbb{E}\right)} =
    \begin{pmatrix}
        X^G_{ii} P_i\mathbb{I} + Y^G_{ii} \mathbb{Q}_i & 
        X^G_{ji} P_j\mathbb{I} + Y^G_{ji} \mathbb{Q}_j \\
        X^G_{ij} P_i\mathbb{I} + Y^G_{ij} \mathbb{Q}_i & 
        X^G_{jj} P_j\mathbb{I} + Y^G_{jj} \mathbb{Q}_j \\
        Y^H_{ii} & Y^H_{ji} \\
        Y^H_{ij} & Y^H_{jj} \\
    \end{pmatrix},
\end{equation}
where $\mathbb{N}=\boldsymbol{n}\otimes\boldsymbol{n}$ is the normal 
projection operator, $\mathbb{T}=\mathbb{I}-\mathbb{N}$ is the 
tangential projection operator (being $\mathbb{I}$ the identity 
matrix) and $\mathbb{N}_{\times}=\boldsymbol{n} \times$ is the 
cross-product operator. 
The ad-hoc defined operators, $P_i\mathbb{I}$ and $\mathbb{Q}_i$, 
can be written as
\begin{equation}
P_i\mathbb{I} = \left( \mathbb{E}^\infty_i \mathbin{:} \mathbb{N} -
\dfrac{1}{3}\mathrm{Tr}\mathbb{E}^\infty_i \right) \mathbb{I},
\end{equation}
\begin{equation}
\mathbb{Q}_i = 2\mathbb{E}^\infty_i -
    2\left(\mathbb{E}^\infty_i \mathbin{:} \mathbb{N} \right)\mathbb{I}.
\end{equation}
Finally, the scalar coefficients $X$ and $Y$ contain the diverging terms
of the lubricant interactions,
\begin{equation}
    X = g^X \dfrac{1}{d_{ij}^{*}+\varepsilon}, \quad \quad 
    Y = g^Y \log \dfrac{1}{d_{ij}^{*}+\varepsilon},
\end{equation}
being $d_{ij}^{*} = 2 d_{ij}/(a_i+a_j)$ the normalised particle-particle
surface distance and $\varepsilon$ is a small positive coefficient that 
avoids the singularity when the particles are at contact.
The coefficients $g^X$ and $g^Y$, instead, are defined as follow:
\begingroup
\allowdisplaybreaks
\begin{align*}
    g^{X^{A}_{ii}}(\alpha)&=2 a_i\dfrac{\alpha^2}{\left(1+\alpha\right)^3}, & 
    g^{X^{A}_{jj}}(\alpha)&=\alpha g^{X^{A}_{ii}}(\alpha^{-1}), \\ 
    g^{X^{A}_{ij}}(\alpha)&=-2 (a_i+a_j)\dfrac{\alpha^2}{\left(1+\alpha\right)^4}, & 
    g^{X^{A}_{ji}}(\alpha)&=g^{X^{A}_{ij}}(\alpha^{-1}), \\ 
    g^{Y^{A}_{ii}}(\alpha)&=\dfrac{4 a_i}{15}
      \dfrac{\alpha\left(2+\alpha+2\alpha^2 \right)}{\left(1+\alpha\right)^3}, &
    g^{Y^{A}_{jj}}(\alpha)&=\alpha g^{Y^{A}_{ii}}(\alpha^{-1}), 
      \tag{\stepcounter{equation}\theequation} \\ 
    g^{Y^{A}_{ij}}(\alpha)&=-\dfrac{4 (a_i+a_j)}{15}
      \dfrac{\alpha \left(2+\alpha+2\alpha^2 \right)}{\left(1+\alpha\right)^4}, &
    g^{Y^{A}_{ji}}(\alpha)&=g^{Y^{A}_{ij}}(\alpha^{-1}), \\ 
    g^{Y^{B}_{ii}}(\alpha)&=-\dfrac{2 a_i^2}{15}
      \dfrac{\alpha\left(4+\alpha \right)}{\left(1+\alpha\right)^2}, &
    g^{Y^{B}_{jj}}(\alpha)&=-\alpha^2 g^{Y^{B}_{ii}}(\alpha^{-1}), \\ 
    g^{Y^{B}_{ij}}(\alpha)&=\dfrac{2 (a_i+a_j)^2}{15}
      \dfrac{\alpha\left(4+\alpha \right)}{\left(1+\alpha\right)^4}, &
    g^{Y^{B}_{ji}}(\alpha)&=-g^{Y^{B}_{ij}}(\alpha^{-1}), \\ 
    g^{Y^{C}_{ii}}(\alpha)&=\dfrac{8 a_i^3}{15}\dfrac{\alpha}{\left(1+\alpha\right)}, &
    g^{Y^{C}_{jj}}(\alpha)&=\alpha^3 g^{Y^{C}_{ii}}(\alpha^{-1}), \\ 
    g^{Y^{C}_{ij}}(\alpha)&=\dfrac{2 (a_i+a_j)^3}{15}
      \dfrac{\alpha^2}{\left(1+\alpha\right)^4}, &
    g^{Y^{C}_{ji}}(\alpha)&=g^{Y^{C}_{ij}}(\alpha^{-1}), \\ 
    g^{X^{G}_{ii}}(\alpha)&=2 a_i^2\dfrac{\alpha^2}{\left(1+\alpha\right)^3}, & 
    g^{X^{G}_{jj}}(\alpha)&=-\alpha^2 g^{X^{G}_{ii}}(\alpha^{-1}), \\ 
    g^{X^{G}_{ij}}(\alpha)&=-2 (a_i+a_j)^2\dfrac{\alpha^2}{\left(1+\alpha\right)^5}, & 
    g^{X^{G}_{ji}}(\alpha)&=-g^{X^{G}_{ij}}(\alpha^{-1}), \\ 
    g^{Y^{G}_{ii}}(\alpha)&=\dfrac{a_i^2}{15}
      \dfrac{\alpha\left(4-\alpha+7\alpha^2 \right)}{\left(1+\alpha\right)^3}, &
    g^{Y^{G}_{jj}}(\alpha)&=-\alpha^2 g^{Y^{G}_{ii}}(\alpha^{-1}), \\ 
    g^{Y^{G}_{ij}}(\alpha)&=-\dfrac{(a_i+a_j)^2}{15}
      \dfrac{\alpha\left(4-\alpha+7\alpha^2 \right)}{\left(1+\alpha\right)^5}, &
    g^{Y^{G}_{ji}}(\alpha)&=-g^{Y^{G}_{ij}}(\alpha^{-1}), \\ 
    g^{Y^{H}_{ii}}(\alpha)&=\dfrac{2 a_i^3}{15}
      \dfrac{\alpha\left(2-\alpha \right)}{\left(1+\alpha\right)^2}, &
    g^{Y^{H}_{jj}}(\alpha)&=\alpha^3 g^{Y^{H}_{ii}}(\alpha^{-1}), \\ 
    g^{Y^{H}_{ij}}(\alpha)&=\frac{(a_i+a_j)^3}{15}
      \dfrac{\alpha^2\left(1+7\alpha \right)}{\left(1+\alpha\right)^5}, &
    g^{Y^{H}_{ji}}(\alpha)&=g^{Y^{G}_{ij}}(\alpha^{-1}), 
\end{align*}
\endgroup
where $\alpha=a_j/a_i$ is the radii ratio.

\section{}\label{app:roll}
In this appendix, we describe the model used to introduce a 
rolling resistance between two contacting particles that can 
be included when necessary.
We follow the work of Luding \citep{LUDING2008} and apply the same algorithm
used for the frictional model,
\begin{subequations}
\begin{align}
    \label{eq:contact1}
    &\boldsymbol{F}^{C,r,*}_{ij} = k_r \boldsymbol{\xi}_{ij}^r, \\
    &\boldsymbol{T}^{C,r}_{ij} = a_{ij} \boldsymbol{n} \times 
                            \boldsymbol{F}^{C,r,*}_{ij},\\
    &\boldsymbol{T}^{C,r}_{ji} = -\boldsymbol{T}^{C,r}_{ij}, 
\end{align}
\end{subequations}
where $a_{ij} = 2(a_i a_j)/(a_i+a_j)$ is the reduced radius. 
The sub- and superscript $r$ state the rolling contribution, 
while the $*$ reports the non-physical nature of the pseudo-force
associated with rolling effects. In fact, $\boldsymbol{F}^{C,r,*}_{ij}$
passively contributes to the rhs of \cref{eq:NE} through the rolling 
torque. The linear stretch $\boldsymbol{\xi}_{r,ij}$ associated to the
rolling spring is computed as in \cref{eq:csi}, with the rolling
velocity $\boldsymbol{u}_r = a_{ij}(\boldsymbol{n} \times
\boldsymbol{\omega_i}-\boldsymbol{n}\times \boldsymbol{\omega_j})$ and 
the rolling spring stiffness $k_r$ (with the constraint $k_r/k_n\ll 1$) 
substituting the relative tangential velocity of the contact points 
$\boldsymbol{u}_t$ and the frictional spring stiffness $k_t$, respectively. 
Note that the limitation introduced by the Coulomb's law is still applied. 
For more details, the reader is referred to \citep{LUDING2008}.

\section*{Data Availability}
The data that support the findings of this study are available from the corresponding author upon reasonable request.

\bibliography{biblio}


\end{document}